\documentclass[12pt]{iopart}
\usepackage{bm}
\usepackage{graphics}
\usepackage{tikz}   %handles figure filenames such as kink_s0.5.eps

\newcommand{\be}{\begin{equation}}
\newcommand{\ee}{\end{equation}}
\newcommand{\ba}{\begin{eqnarray}}
\newcommand{\ea}{\end{eqnarray}}
\newcommand{\pard}[2]{\frac{\partial #1}{\partial #2}}
\newcommand{\vc}[1]{{\bm #1}}
\newcommand{\Bgav}[1]{\Big \langle #1 \Big \rangle}
\newcommand{\gav}[1]{\langle #1 \rangle}
\newcommand{\df}{{\rm d}}

\newcommand{\Ah}{A_{\|}^{{\rm(h)}}}
\newcommand{\As}{A_{\|}^{{\rm(s)}}}

\newcommand{\Sanchez}[1]{{\color{black}#1}}
\newcommand{\Brunner}[1]{{\color{black}#1}}
\begin{document}

\title[Numerics of electromagnetic turbulence]{Numerics and computation in gyrokinetic simulations of electromagnetic turbulence with global particle-in-cell codes}

\author{A. Mishchenko$^{(1)}$, A. Biancalani$^{(2)}$, A. Bottino$^{(2)}$, T.~Hayward-Schneider$^{(2)}$,
Ph. Lauber$^{(2)}$, E.~Lanti$^{(4)}$, L.~Villard$^{(3)}$, R.~Kleiber$^{(1)}$, A.~{K\"onies}$^{(1)}$, and M.~Borchardt$^{(1)}$}

\address{$^{(1)}$Max Planck Institute for Plasma Physics, D-17491 Greifswald, Germany\\
$^{(2)}$Max Planck Institute for Plasma Physics, D-85748 Garching, Germany \\
$^{(3)}$Ecole Polytechnique F\'ed\'erale de Lausanne (EPFL), Swiss Plasma Center (SPC), CH-1015 Lausanne, Switzerland\\
$^{(4)}$ Ecole Polytechnique F\'ed\'erale de Lausanne (EPFL), SCITAS, CH-1015 Lausanne, Switzerland}
\ead{alexey.mishchenko@ipp.mpg.de}
\vspace{10pt}
\begin{indented}
\item[]March 2021
\end{indented}

\begin{abstract}
Electromagnetic turbulence is addressed in tokamak and stellarator plasmas with the global gyrokinetic particle-in-cell codes ORB5 [E.~Lanti~et~al, Comp.~Phys.~Comm, {\bf 251}, 107072 (2020)] and EUTERPE \Sanchez{[V.~Kornilov~et~al, Phys.~Plasmas, {\bf 11}, 3196 (2004)]}. The large-aspect-ratio tokamak, down-scaled ITER, and Wendelstein 7-X geometries are considered. The main goal is to increase the plasma beta, the machine size, the ion-to-electron mass ratio, as well as to include realistic-geometry features in such simulations. The associated numerical requirements and the computational cost for the cases on computer systems with massive GPU deployments are investigated. These are necessary steps to enable electromagnetic turbulence simulations in future reactor plasmas.
\end{abstract}

\noindent{\it Keywords}: gyrokinetics, particle-in-cell, turbulence
%\maketitle
\section{Introduction}

Realistic simulations of electromagnetic turbulence are of crucial importance to understand and predict behaviour of burning plasmas before such plasmas become experimentally available. Burning plasmas are complex systems with multiple spatial and temporal scales \cite{Zonca}.
%A substantial energetic-particle minority couples electromagnetic turbulence, global Alfv\'enic and Magneto-Hydrodynamical (MHD) modes, and zonal flows.
%
Electromagnetic turbulence is ubiquitous in such plasmas. It is the basic component of the ``scenery'' involving the fast-particle dynamics, the global Magneto-Hydrodynamical (MHD) and Alfv\'enic activity, zonal flows, and transport. The saturation of the electromagnetic turbulence is a consequence of a complex interplay between these components.
A single numerical first-principle framework, based on the global gyrokinetic electromagnetic formulation and including self-consistently all of the parts of the problem, is desirable.
%Addressing this physics in its full complexity shall become possible on future exascale platforms. Enabling the codes to run on the existing and emerging heterogeneous computer systems is a crucial step in the preparation to the exascale era.

\Sanchez{
There are a number of gyrokinetic codes which develop capabilities for global simulations in electromagnetic regimes, for example GEM \cite{GEM_reference}, GTC \cite{Holod_GTC,Dong_GTC}, XGC \cite{XGC_pullback}, GENE \cite{GENE_reference}, GKNET \cite{Ishizawa}, GKW \cite{GKW_reference}. In this paper, we employ the well-established gyrokinetic particle-in-cell (PIC) codes ORB5 \cite{Lanti} and EUTERPE \Sanchez{\cite{Kornilov04}}.
The ORB5 code has recently been refactored to allow simulations on heterogeneous (GPU) systems. A very good scaling of the code has been reported in Ref.~\cite{Ohana} on various GPU systems and including electromagnetic perturbations. Our paper extends these results to massive production simulations of electromagnetic turbulence deploying GPUs on the Marconi100 computing system (CINECA). We believe that it is a step forward in the adoption of this new HPC technology by the gyrokinetic PIC simulation community.
Electromagnetic simulations are known to be very challenging for the gyrokinetic particle-in-cell codes because of the numerical stability issues related to the cancellation problem \cite{YangChen,Mishchenko_mitigation}. Such simulations are also very time consuming since the fast electron dynamics has to be resolved. In ORB5 and EUTERPE, we address the numerical stability problem using the pullback mitigation technique \cite{pullback,ORB5_pullback} for the cancellation problem. Very long simulation times normally required when the electron dynamics is resolved are substantially accelerated in the case of ORB5 on the GPUs of the Marconi100 computer system. This paper reports on the first massive usage of the GPU accelerators in ORB5 production runs.

Electromagnetic turbulence has already been addressed using ORB5 in Ref.~\cite{Biancalani}. While being a groundbreaking effort, this work had a number of limitations. For example, the plasma beta $\beta = 0.1\%$ and the machine size $L_x = 2 r_a/\rho_i = 350$ were quite small. Here, $r_a$ is the minor radius of the tokamak and $\rho_i$ is the characteristics ion gyro-radius. Also, the geometry was limited to a large-aspect-ratio circular cross-section tokamak and the typical mass ratio was reduced to $m_i/m_e = 200$ (although a limited number of simulations using larger mass ratios were also performed). However, it is known \cite{YangChen,Mishchenko_mitigation} that the numerical error caused by the cancellation problem scales as $\delta A_{\|} \sim L_x^2 \, \beta \, m_i/m_e$. Also, the more complex shaping of the magnetic surfaces in realistic tokamak and stellarator geometries can complicate the cancellation problem \cite{Mishchenko_shaping}. Thus, it still remains to be proven that the gyrokinetic particle-in-cell simulations of electromagnetic turbulence are possible under more challenging, realistic, conditions. This demonstration is the main purpose of our present paper. In the following, we will show that simulations of electromagnetic turbulence using more realistic parameters and in more complex geometries are possible with ORB5 \cite{Lanti} and EUTERPE \cite{Kornilov04} in their present form and on currently available computer systems. We have chosen ITER and W7-X geometries for the first assessment of the electromagnetic turbulence in the shaped plasmas using the global gyrokinetic particle-in-cell codes.

In the finite-beta regimes, considered in this paper, the plasma experiences a transition from the Ion Temperature Gradient driven (ITG) to the Kinetic Ballooning Mode (KBM) turbulence which becomes dominant when beta increases. An important result of this work is the demonstration of numerically stable global particle-in-cell simulations of the electromagnetic turbulence in the KBM regime. This is a novel result for global fully-gyrokinetic PIC codes. Previously, the global simulations of the ballooning and KBM-type turbulence have been performed using MHD \cite{Zhu_Hegna_NIMROD} and two-fluid \cite{BOUT_Ma_2016} codes. Most of the gyrokinetic studies \cite{Pueschel_GENE,Waltz_GYRO,Terry_2015,Di_Siena_2019} have used the flux-tube approximation and Eulerian discretization of the nonlinear gyrokinetic system. There are only two examples of global gyrokinetic KBM turbulence simulations, reported so far: Ref.~\cite{Ishizawa} using the Eulerian code GKNET and Ref.~\cite{Dong_GTC} where the gyrokinetic PIC code GTC has been used. The principle difference of Ref.~\cite{Dong_GTC} comparing to our paper is the treatment of the electrons. In Ref.~\cite{Dong_GTC}, the electrons are treated in the fluid approximation ("fluid-kinetic hybrid electron model") whereas this papers uses the fully gyrokinetic description for all species. In the fluid-electron approximation, the cancellation problem is not an issue, in contrast to our simulations where the cancellation problem must be solved.

The novel simulations reported in this paper have been made possible by a combination of several recent developments. First, the pullback mitigation of the cancellation problem implemented in EUTERPE \cite{pullback} and more recently in ORB5 \cite{ORB5_pullback} has stabilized numerically electromagnetic simulations and relaxed considerably the resolution requirement for such simulations. Second, the GPU enabling of ORB5 \cite{Ohana} made it possible to run the large number of small time steps, needed to resolve the electron dynamics, in a reasonable simulation time since the markers can now be pushed on the GPUs. The third aspect, making this paper possible, is a large amount of computing time available for ORB5 and EUTERPE simulations in the frame of PRACE 21st Call for Project Acces. In particular, all ORB5 simulations reported in this paper have been perofmed on the GPU computing systems Marconi100 installed very recently at CINECA using the computing time from the PRACE project.
In this paper, we will not be able to clarify every aspect of the global electromagnetic turbulence physics which is a quite complex topic at the frontier of the ongoing research. However, we hope to contribute to this research providing an insight into the details of the simulations which may help future progress.
}

The structure of this paper is as follows. In Sec.~\ref{Equations}, we present, for completeness, the equations solved by the code. In Sec.~\ref{Simulations}, simulations are discussed for the large-aspect-ratio circular cross-section tokamak \cite{Biancalani}, for the down-scaled ITER \cite{Hayward}, and for the down-scaled Wendelstein 7-X (W7-X) configurations. Conclusions are made in Sec.~\ref{Conclusions}.
%
%=================================================================
%
\section{Equations and discretization} \label{Equations}
In this section, we present for completeness equations solved by the codes. We use the mixed-variable formulation of gyrokinetic theory. This formulation is based on the splitting of the magnetic potential into the Hamiltonian and symplectic parts $A_{\|} = A_{\|}^{{\rm(h)}} + A_{\|}^{{\rm(s)}}$, see Ref.~\cite{pullback} for details.
%The pullback mitigation \cite{pullback} is applied for the cancellation problem.

The equations include the gyrokinetic Vlasov equation:
\be
\label{vlasov_lin}
\pard{f_{1s}}{t} + \dot{\vc{R}} \cdot \pard{f_{1s}}{\vc{R}} +
\dot{v}_{\|} \pard{f_{1s}}{v_{\|}} =
{}- \dot{\vc{R}}^{(1)} \cdot \pard{F_{0s}}{\vc{R}} -
\dot{v}_{\|}^{(1)} \pard{F_{0s}}{v_{\|}} \ ,
\ee
the equations for the gyro-center orbits:
  \ba
  &&{} \dot{\vc{R}} = \dot{\vc{R}}^{(0)} + \dot{\vc{R}}^{(1)} \ , \;\;\;\;
  \dot{v}_{\|} = \dot{v}_{\|}^{(0)} + \dot{v}_{\|}^{(1)}   \\
  &&{}\dot{\vc{R}}^{(0)} = v_{\|} \vc{b}^*_0 + \frac{1}{q B_{\|}^*}\vc{b}\times \mu \nabla B \ , \;\;\;
  \dot{v}_{\|}^{(0)} = \,-\,\frac{\mu}{m} \,\vc{b}^*_0 \cdot \nabla B
  \\
  \label{dotR1}
  &&{} \dot{\vc{R}}^{(1)} = \frac{\vc{b}}{B_{\|}^*} \times \nabla \Bgav{ \phi -
    v_{\|} A_{\|}^{\rm(s)} - v_{\|} A_{\|}^{\rm(h)} } -
  \frac{q}{m} \,\gav{A^{\rm(h)}_{\|}} \, \vc{b}^*_0 \\
  \label{dotp1}
  &&{} \dot{v}_{\|}^{(1)} = \,-\, \frac{q}{m} \,
  \left[ \vc{b}^* \cdot \nabla \Bgav{\phi - v_{\|} A_{\|}^{\rm(h)}} + \pard{}{t}
    \Bgav{A_{\|}^{\rm(s)}} \right]
   -  \frac{\mu}{m} \, \frac{\vc{b} \times \nabla B}{B_{\|}^*} \cdot \nabla
    \Bgav{A_{\|}^{\rm(s)}} \\
  \label{bstar}
  &&{}\vc{b}^* = \vc{b}^*_0 + \frac{\nabla\gav{A_{\|}^{{\rm(s)}}}\times \vc{b}}{B_{\|}^*} \ , \;\;\; \vc{b}^*_0 = \vc{b} + \frac{m v_{\|}}{q B_{\|}^*} \nabla\times\vc{b} \\
  &&{} B_{\|}^* = B + \frac{m v_{\|}}{q} \vc{b}\cdot\nabla\times\vc{b} \ ,
  \ea
  an equation for $\partial A_{\|}^{\rm(s)} / \partial t$ (Ohm's law, see \cite{pullback}):
  \be
  \label{Ohm}
  \pard{}{t}A_{\|}^{\rm(s)} + \vc{b} \cdot \nabla \phi = 0 \ ,
  \ee
  %
  %The zeroth-order gyrocenter characteristics remain unchanged.
  %The mixed-variable distribution function is solved from the gyrokinetic Vlasov
  %equation.
  %
  and the field equations (here we use the notation as in Ref.~\cite{pullback}):
  \ba
  \label{qasi}
  &&{}
  \,-\,\nabla\cdot\left(\frac{n_0}{B \omega_{ci}}\nabla_{\perp}\phi\right)
   = \bar{n}_{1i} - \bar{n}_{1e} \\
  \label{amp}
  &&{}
  \sum_{s = i,e}\frac{\beta_s}{\rho_{s}^{2}} A_{\|}^{\rm(h)}
  - \nabla_{\perp}^{2} A_{\|}^{\rm(h)}
  = \mu_{0} \sum_{s = i,e} \bar{j}_{\|1s} + \nabla_{\perp}^2 A_{\|}^{\rm(s)}
  \ea
%For the derivation and further details, see Ref. \cite{pullback}.
  %
%Here, the standard notation is used.
For the cancellation problem, the pullback mitigation \cite{pullback} is applied.
\begin{enumerate}
\item At the end of each time step, the splitting of the magnetic potential is redefined:
\be
\label{ic_A}
A_{\|\rm(new)}^{\rm(s)}(t_i) = A_{\|}(t_i) = A_{\|\rm(old)}^{\rm(s)}(t_i) +
A_{\|\rm(old)}^{\rm(h)}(t_i)
%\longrightarrow %\ , \;\;\; A_{\|}^{\rm(h)}(t_i) \longrightarrow 0
\ee
\item As a consequence, $A_{\|\rm(new)}^{\rm(h)}(t_i) = 0$
\item The new mixed-variable distribution function and the parallel velocity become \cite{Kleiber_pullback}:
%\be
%f_{s}(Z_s,  A_{\|}^{(s)}) = f_{m}(Z_m,  A_{\|}^{(s)}, A_{\|}^{(h)})
%\ee
%For the non-linear pullback, we transform the pertubed distribution and the parallel velocity, see Ref.~\cite{Kleiber_pullback} for details: %[R.~Kleiber~et~al, PoP 2016]:
\ba
\label{ic_f_nlin}
&&{} f_{1s{\rm(new)}}(v_{\|}^{\rm(s)},  A_{\|}^{{\rm(s)}}) = f_{1s{\rm(old)}}(v_{\|}^{\rm(m)},  A_{\|}^{{\rm(s)}}, A_{\|}^{{\rm(h)}}) + F_{0}(v_{\|}^{\rm(m)}) - F_{0}(v_{\|}^{\rm(s)})
\\
&&{} v_{\|}^{\rm(s)} = v_{\|}^{\rm(m)} %+ \frac{\vc{b}^*}{B} \cdot \nabla \left[
%\intl^{\theta_{\rm(gc)}} \Big( \psi - \gav{\psi} \Big) \, \df \theta_{\rm(gc)} \right]
- \frac{e}{m} \, \Bgav{A_{\|}^{\rm(h)}}
\ea
The linearized version of the pullback transformation is:
\be
\label{ic_f_lin}
f_{1s\rm(new)}(t_i) = %f_{1s}^{\rm(s)}(t_i) =
f_{1s\rm(old)}(t_i) + \frac{q_s \, \gav{A_{\|\rm(old)}^{\rm(h)}(t_i)}}{m_s}  \,
\pard{F_{0s}}{v_{\|}}
%= f_{1s} - \frac{q_s}{T_s} \, \Big[ v_{\|} - u_{\|s0}
%\Big] \gav{A_{\|}^{\rm(h)}} \, F_{0s}
\ee
The velocity is not transformed when the linearized pullback is applied and the approximation $\vc{b}^* = \vc{b}^*_0$ is used, see Eq.~(\ref{bstar}).
\item The nonlinear mixed-variable system of equations
  (\ref{dotR1})-\Sanchez{(\ref{amp})} is explicitly solved
  at the next time step $t_i + \Delta t$ in a usual
  way, but using the transformed coordinates and the transformed distribution function as the initial conditions.
%
%\item Perform this transformation at  the end of each time step.
%
\end{enumerate}
%(symplectic-hamiltonian equivalence at the 2nd order)
\Sanchez{
Details of the pullback implementation have been described in Ref.~\cite{ORB5_pullback} for ORB5 and in Ref.~\cite{pullback} for EUTERPE. The nonlinear pullback transformation has been introduced in Ref.~\cite{Kleiber_pullback} using the GYGLES code.
The nonlinear equations of motion solved in GYGLES and described in Ref.~\cite{Kleiber_pullback} are considerably more complicated than the equations implemented in ORB5 \cite{ORB5_pullback} and EUTERPE \cite{pullback} where the nonlinear terms proportional to $\Ah$ (e.~g.~the terms proportional to $\Bgav{\Ah}^2$ or $\Bgav{\Ah}\Bgav{\As}$) have been omitted for simplicity. It has been seen numerically (using GYGLES) that these terms do not affect the nonlinear results. The reason for that is the smallness of $\Ah$ which is set to zero at the end of every time step. Another approximation used in ORB5 is $\nabla \times (\As \vc{b}) \approx \nabla \As \times \vc{b}$ which amounts for a small geometric correction normally unimportant. These small omitted terms may however play certain role at longer simulation times. This subject may need further assessment in future. In this present paper, we limit our formulation to keeping the dominant nonlinear terms, proportional to $\vc{b}\times\nabla\Bgav{\As} \cdot \nabla \gav{\phi}$ and corresponding to the nonlinear magnetic flutter. This nonlinearity arises in the equations of motion from the perturbed term kept in $\vc{b}^*$, Eq.~(\ref{bstar}). In the next Section, we show that these larger nonlinear terms are of crucial importance for the turbulence saturation at a higher beta.
}
%
%=================================================================
%
\section{Simulations of electromagnetic turbulence}  \label{Simulations}
\subsection{Increasing plasma beta}
We consider geometry and parameters described in Ref.~\cite{Biancalani}: a tokamak with aspect ratio 10, machine size $L_x = 350$, circular cross-section, the safety factor $q(\rho) = 1.8503 - 0.074096\rho + 0.95318 \rho^2 - 5.703 \rho^3 + 7.5779 \rho^4 - 0.19182 \rho^5 - 1.8253 \rho^6$ (here, $\rho$ is the minor radius of the circular flux surfaces), and the plasma profiles: %given by the expressions:
\ba
\label{dens_prof}
&&{} n_{0s}(s)/n_{0s}(s_0) = \exp\left[-\kappa_n \Delta_n {\rm tanh}\left(\frac{s - s_0}{\Delta_n}\right)\right] \\
\label{temp_prof}
&&{} T_{0s}(s)/T_{0s}(s_0) = \exp\left[-\kappa_T \Delta_T {\rm tanh}\left(\frac{s - s_0}{\Delta_T}\right)\right]
\ea
with $s = \sqrt{\psi/\psi_a}$, $\psi$ the poloidal magnetic flux, $\psi_a$ the poloidal magnetic flux at the plasma edge, $s_0 = 0.5$, $\kappa_n = 0.3$, $\kappa_T = 1.0$, and $\Delta_T = \Delta_n = 0.208$. The ubiquitous nonlinear generation of the small scales in the phase space (filamentation) is controlled with the Krook operator.
\Sanchez{We have used the standard noise-correction Krook operator described in Ref.~\cite{McMillan2008} for ORB5 and further detailed in Ref.~\cite{Sanchez_EUTERPE} for EUTERPE. Following \cite{Biancalani}, $\gamma_{Krook} = 10^{-4}\omega_{ci}$ is applied for the ions and for the electrons with only the zonal-flow conserving correction applied in ORB5 simulations, see Ref.~\cite{McMillan2008} for further implementation details.} %and the associated damping rate $\gamma = 10^{-4} \omega_{ci}$, see Ref.~\cite{Biancalani} for further details.
 Keeping all these parameters unchanged, we increase the plasma beta from $\beta = 0.1\%$ to a more realistic $\beta = 1.6\%$. In contrast to Ref.~\cite{Biancalani}, we use the nonlinear pullback transformation and do not include the fast particles.
  %In Ref.~\cite{Biancalani}, the linear pullback transformation, Eq.~(\ref{ic_f_nlin}), has been used together with the approximation $\vc{b}^* =  \vc{b}^*_0$. It has been shown that these approximations have a marginal effect on the results in the low-beta ($\beta = 0.1\%$) regime. One can show that using the nonlinear pullback is mandatory for higher $\beta$. Otherwise, a numerical instability develops in a nonlinear phase.

\begin{figure}
\includegraphics[width=0.47\textwidth]{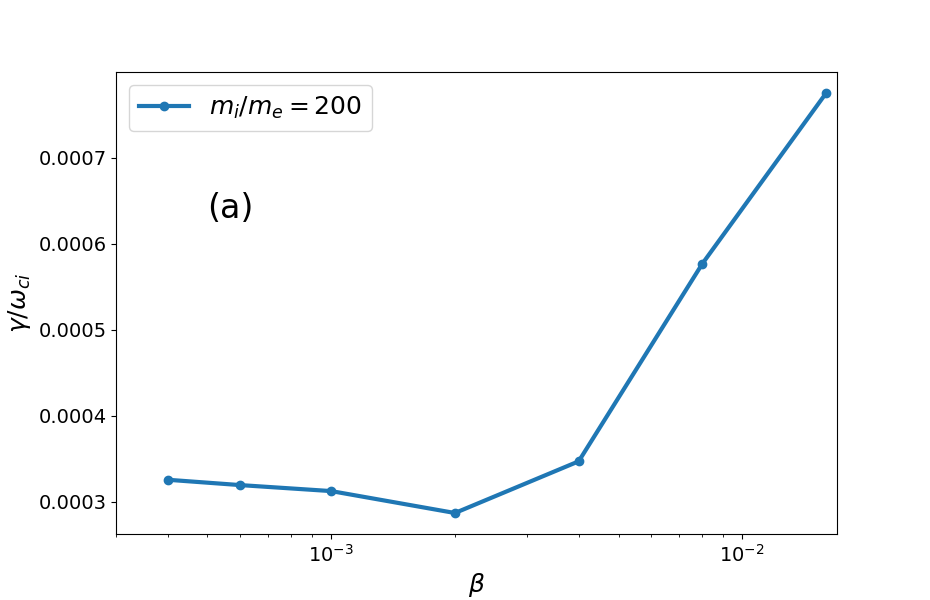} %{{./graph/gamma_beta_ITG2KBM.png}}
\includegraphics[width=0.47\textwidth]{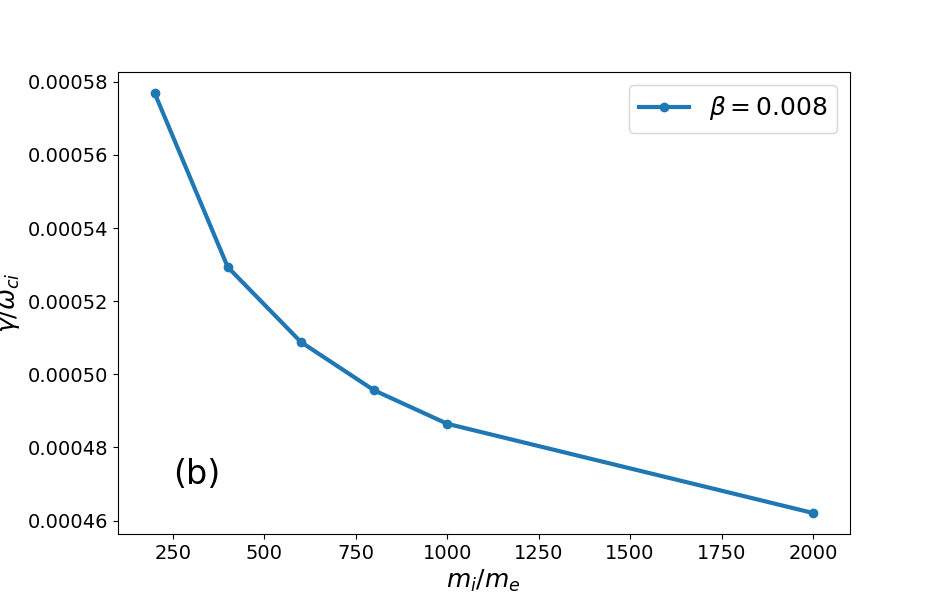} %{{./graph/gamma_me-b0.002.png}}
\caption{\Brunner{
(a) The growth rate (normalized to $\omega_{ci}$) shown as a function of $\beta$ for $m_i/m_e = 200$. One can see a typical ITG-to-KBM transition pattern \cite{Ishizawa,XGC_pullback}. (b) The growth rate shown as a function of the mass ratio for $\beta = 0.8\%$. The growth rate is computed from the linear phase of the perturbed energy flux evolution. The Krook decay rate $\gamma_{Krook}/\omega_{ci} = 10^{-4}$, used here, is not much smaller than the growth rate.
}}
\label{gamma_eflux}
\end{figure}
\Brunner{In Fig.~\ref{gamma_eflux}(a), the growth rate normalized to the ion cyclotron frequency $\omega_{ci}$ is shown as a function of $\beta$ for $m_i/m_e = 200$. One can see how the growth rate decreases with beta until a certain threshold is reached beyond which the growth rate begins to strongly increase. This is a typical ITG-to-KBM transition pattern \cite{Ishizawa,XGC_pullback}. Here, the growth rate is computed from the linear phase of the perturbed energy flux evolution, see below for more details. In Fig.~\ref{gamma_eflux}(b), the growth rate is plotted for different mass ratios including the realistic one for $\beta = 0.8\%$. One sees that the system is somewhat stabilized when the mass ratio increases. One sees that $\gamma_{Krook}/\omega_{ci} = 10^{-4}$ is not much smaller than the growth rate shown in Fig.~\ref{gamma_eflux}. We correct it to $\gamma_{Krook} = \gamma_{lin}/10$ in the following.}
\begin{figure}
\includegraphics[width=0.47\textwidth]{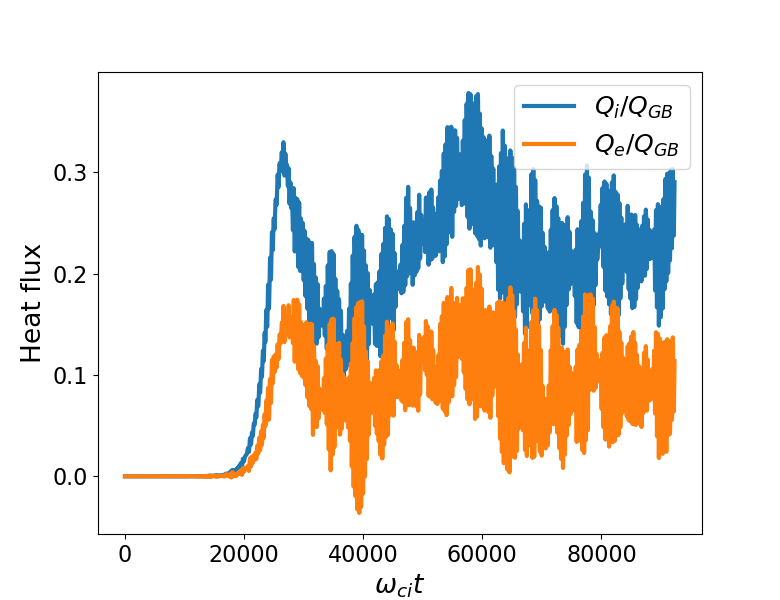} %{{./graph/hflux_t_b0.00025.png}}
\includegraphics[width=0.47\textwidth]{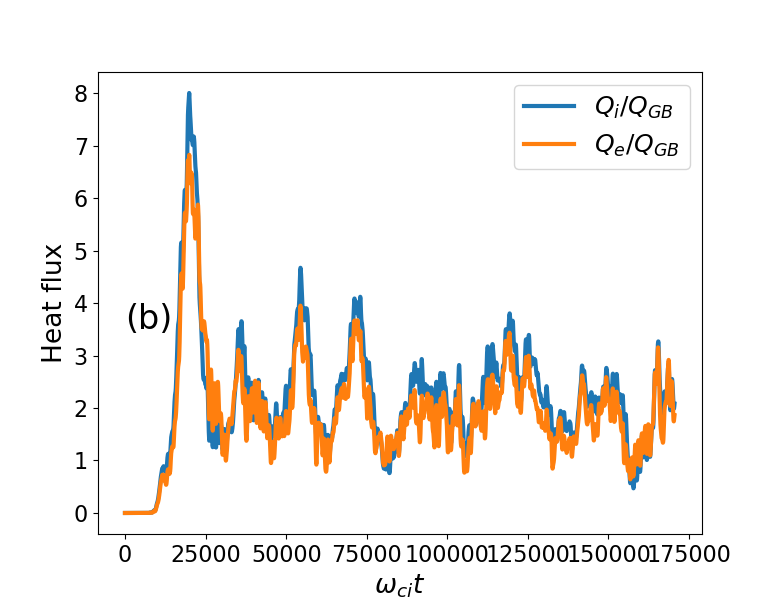} %{{./graph/hflux_t_b0.004.png}}
\caption{\Brunner{
The averaged radial heat flux density in gyro-Bohm units for (a) $\beta = 1.6\%$ (ITG regime) and (b) $\beta = 1.6\%$ (KBM regime). The heat flux is considerably larger in the KBM regime.
}}
\label{eflux_NlinPB}
\end{figure}

\Brunner{In Fig.~\ref{eflux_NlinPB}, the radial heat flux evolution is plotted for (a) $\beta = 0.1\%$ (ITG regime) and (b) $\beta = 1.6\%$ (KBM regime). The time is normalized to the ion cyclotron frequency $\omega_{ci}$. The radial heat flux density is normalized to the gyro-Bohm units $Q_{GB} = n_e T_e c_s (\rho_s/a)^2$ with the sound speed $c_s^2 = T_e/m_i$ and the ion sound gyro-radius $\rho_s = c_s/\omega_{ci}$. To compute the fluxes and the perturbed profiles, the plasma volume is divided into $N_b$ radial bins, each of volume $V_j , \; j \in [1,N_b]$. The heat flux density $Q^{(j)}_s = Q^{(j)}_{0s} + Q^{(j)}_{1s} - (5/2) T_s^{(j)} P^{(j)}_{1s}$ is computed using the expressions:
\ba
Q^{(j)}_{0s} =
\frac{1}{V_j} \int\limits_{V_j}\df\vec{R}\int F_{0s} \frac{m_s v^2}{2} \frac{\dot{\vc{R}}^{(1)} \cdot\nabla\psi}{|\nabla\psi|} \df\vc{v} \\
Q^{(j)}_{1s} =
\frac{1}{V_j} \int\limits_{V_j}\df\vec{R}\int f_{1s} \frac{m_s v^2}{2} \frac{\dot{\vc{R}}^{(1)} \cdot\nabla\psi}{|\nabla\psi|} \df\vc{v} \\
P^{(j)}_{1s} =
\frac{1}{V_j} \int\limits_{V_j}\df\vec{R}\int f_{1s} \frac{\dot{\vc{R}}^{(1)} \cdot\nabla\psi}{|\nabla\psi|} \df\vc{v}
\ea
The mean value of $Q^{(j)}_s$ is plotted in Fig.~\ref{eflux_NlinPB}. Note that this is the total radial heat flux density with the convective part subtracted. The growth rate shown in Fig.~\ref{gamma_eflux} is computed using $Q^{(j)}_{1s}$ only which is assumed to evolve as $Q^{(j)}_{1s} \sim \exp(2\gamma t)$ during the linear phase on the evolution. The evolving plasma temperature and density are computed using the expressions:
\ba
\label{temp_def}
&&{} T^{(j)}_s = \frac{1}{3} m_s \left(\{v_{\perp}^2\}^{(j)}_s + \{v_{\|}^2\}^{(j)}_s - \{v_{\|}\}^{(j)}_s \{v_{\|}\}^{(j)}_s\right) \\
\label{dens_def}
&&{} n_{sj} = \frac{1}{V_j} \int\limits_{V_j}\df\vec{R}\int f_s \df\vc{v} \ , \;\; f_s = F_{0s}+f_{1s} \\
&&{} \{v_{\perp}^2\}^{(j)}_s = \frac{1}{n_{sj} V_j} \int\limits_{V_j}\df\vec{R}\int f_s v_{\perp}^2 \df\vc{v} \ , \;\;
\{v_{\|}^2\}^{(j)}_s = \frac{1}{n_{sj} V_j} \int\limits_{V_j}\df\vec{R}\int f_s v_{\|}^2 \df\vc{v}\\
&&{} \{v_{\|}\}^{(j)}_s = \frac{1}{n_{sj} V_j} \int\limits_{V_j}\df\vec{R}\int f_s v_{\|} \df\vc{v}
\ea
}

\begin{figure}
\includegraphics[width=0.47\textwidth]{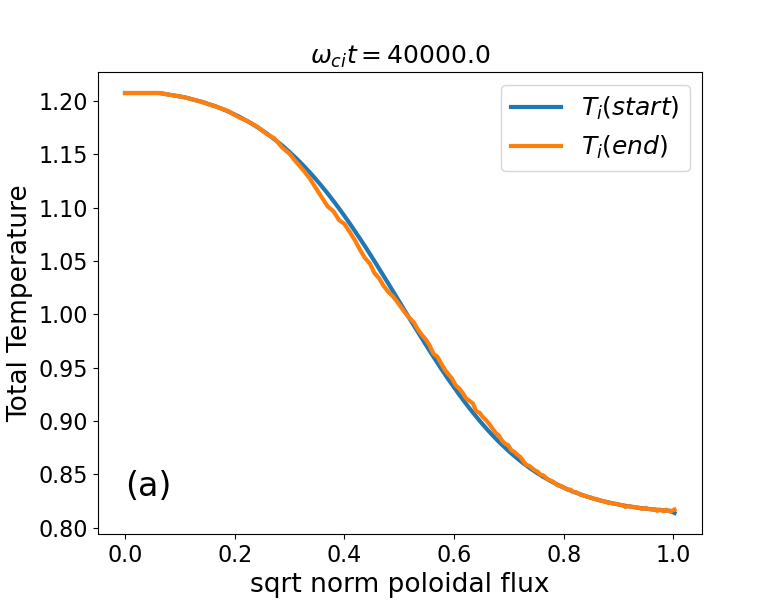} %{{./graph/temp_relaxed_b0.00025.png}}
\includegraphics[width=0.47\textwidth]{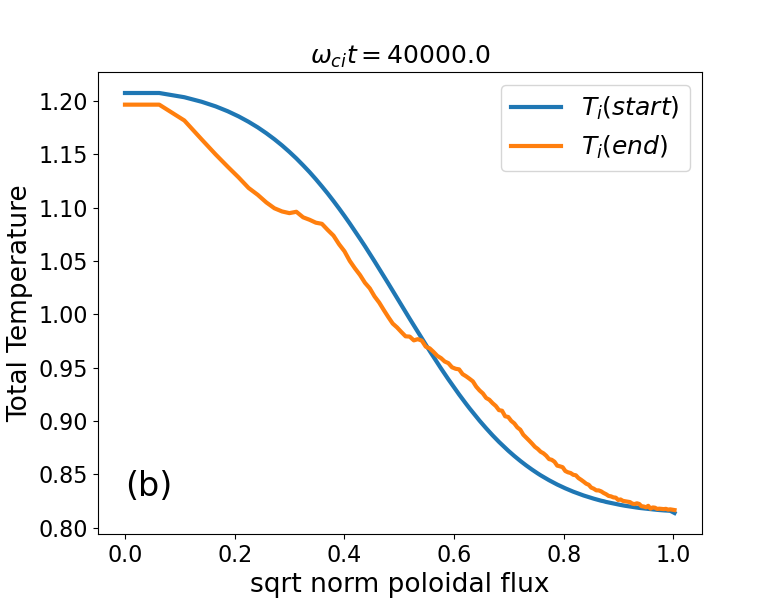} %{{./graph/temp_relaxed_b0.004.png}}
\caption{\Brunner{
Evolution of the total temperature profile for (a) $\beta = 0.1\%$ (ITG regime) and (b) $\beta = 1.6\%$ (KBM regime). The temperature is computed using Eq.~(\ref{temp_def}). The temperature relaxation is much stronger in the KBM case.
}}
\label{temp_BAE}
\end{figure}
\begin{figure}
\includegraphics[width=0.47\textwidth]{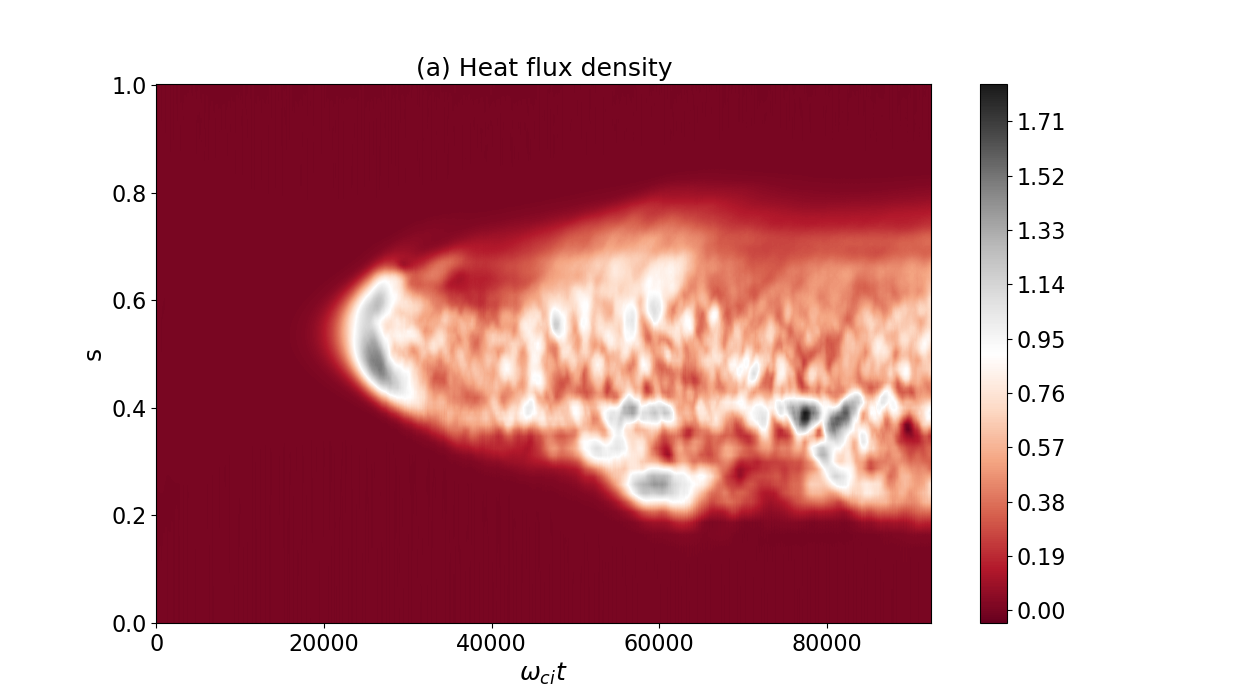} %{{./graph/eflux_st_b0.00025.png}}
\includegraphics[width=0.47\textwidth]{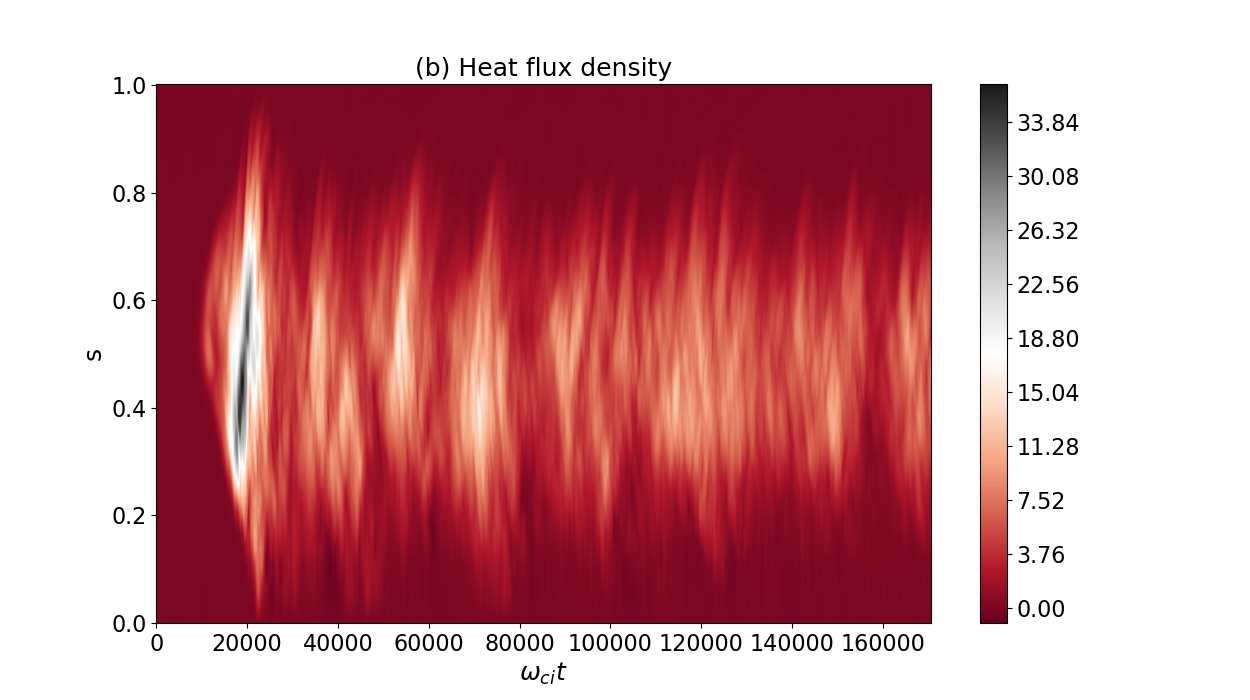} %{{./graph/eflux_st_b0.004.png}}
\caption{\Brunner{
The radial heat flux evolution in the gyro-Bohm units shown for (a) $\beta = 0.1\%$ (ITG regime) and (b) $\beta = 1.6\%$ (KBM regime). The heat flux is radially more localized in the ITG case. In the KBM regime, the turbulence evolves as a sequence of broad relaxation events.
}}
\label{eflux_st_BAE}
\end{figure}
\begin{figure}
\includegraphics[width=0.47\textwidth]{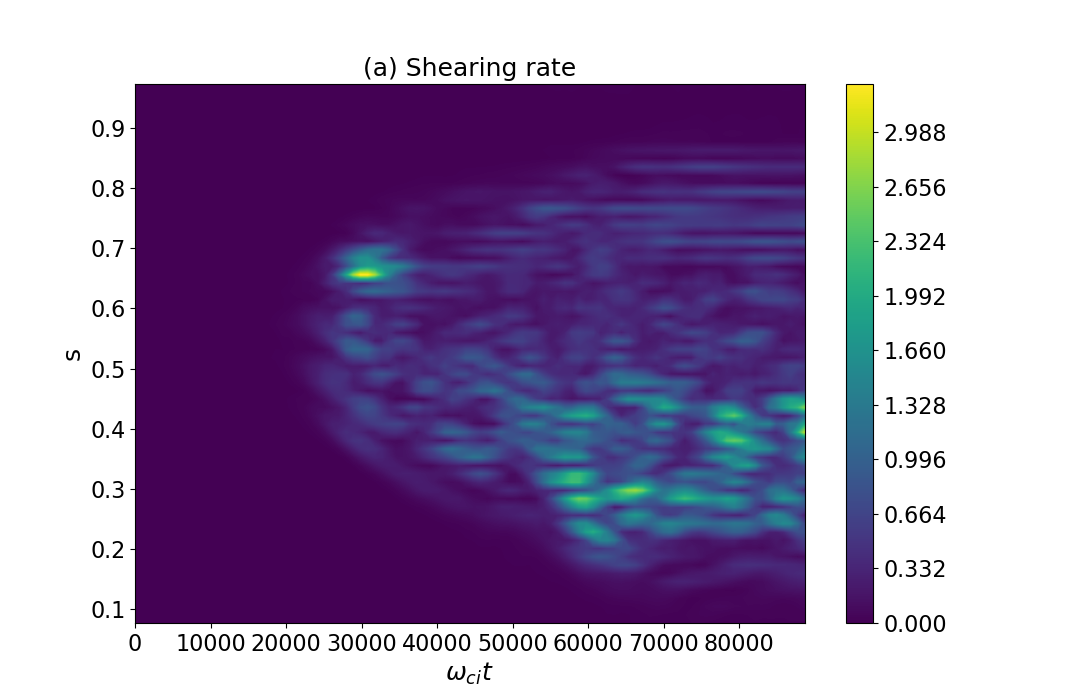} %{{./graph/shearing_rate_b0.00025.png}}
\includegraphics[width=0.47\textwidth]{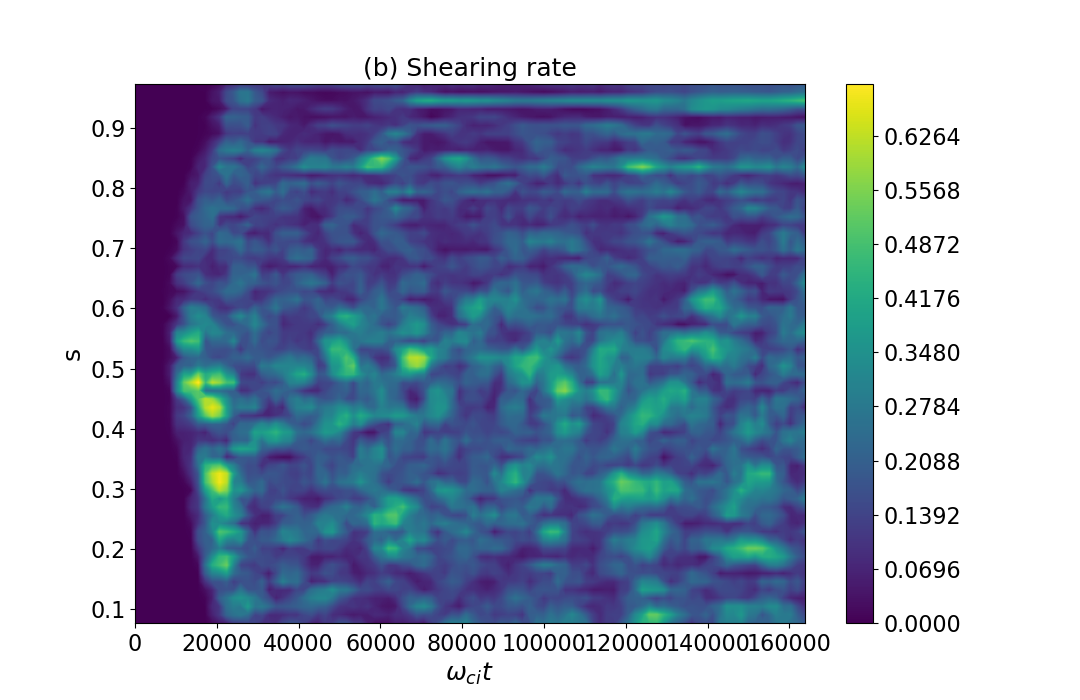} %{{./graph/shearing_rate_b0.004.png}}
\caption{\Brunner{
The zonal-flow shearing rate normalized to $\gamma_{lin}$ is shown for (a) $\beta = 0.1\%$ (ITG regime) and (b) $\beta = 1.6\%$ (KBM regime). The normalized shearing rate is smaller in the KBM regime.
}}
\label{shearing_BAE}
\end{figure}

\Brunner{
In Fig.~\ref{temp_BAE}, the temperature relaxation is shown for (a) the ITG regime and (b) the KBM regime. The temperature is computed using Eq.~(\ref{temp_def}). One sees that the temperature relaxes much stronger in the KBM case. Accordingly, the ion heat flux evolution, plotted in Fig.~\ref{eflux_st_BAE}(b), shows a sequence of the broad relaxation events in the KBM case whereas the turbulence is more localized around the unperturbed temperature gradient in the ITG case, Fig.~\ref{eflux_st_BAE}(a). This localization is a consequence of the weaker temperature relaxation for the ITG turbulence observed in Fig.~\ref{eflux_st_BAE}(a). This broader radial distribution of the KBM turbulence is also reflected in the shearing rate roughly estimated here as $\alpha_E \approx (1/L_x)^2 \partial^2 \phi_{00}/ \partial s^2$ with $\phi_{00}$ the "zonal" component of the electrostatic potential. The shearing rate is shown in Fig.~\ref{shearing_BAE} for (a) the ITG regime and (b) the KBM regime. We normalize the shearing rate to the corresponding linear growth rate $\gamma_{lin}$ shown in Fig.~\ref{gamma_eflux}(a). One can see that the distribution of $\alpha_E$ is much broader in the KBM case although its value is smaller than in the ITG case.
}

\begin{figure}
\includegraphics[width=0.33\textwidth]{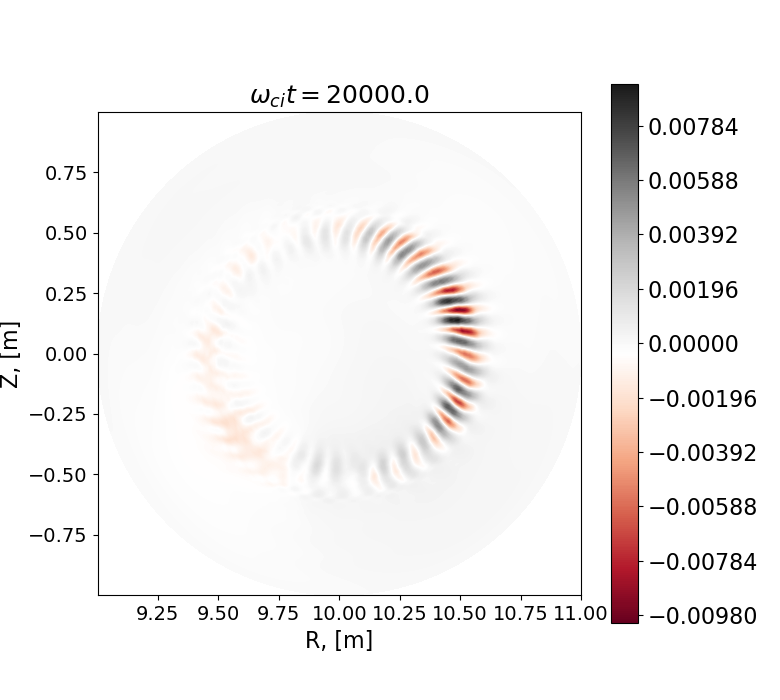} %{{./graph/BAE_b0.00025/potsc_t20e3.png}}
\includegraphics[width=0.33\textwidth]{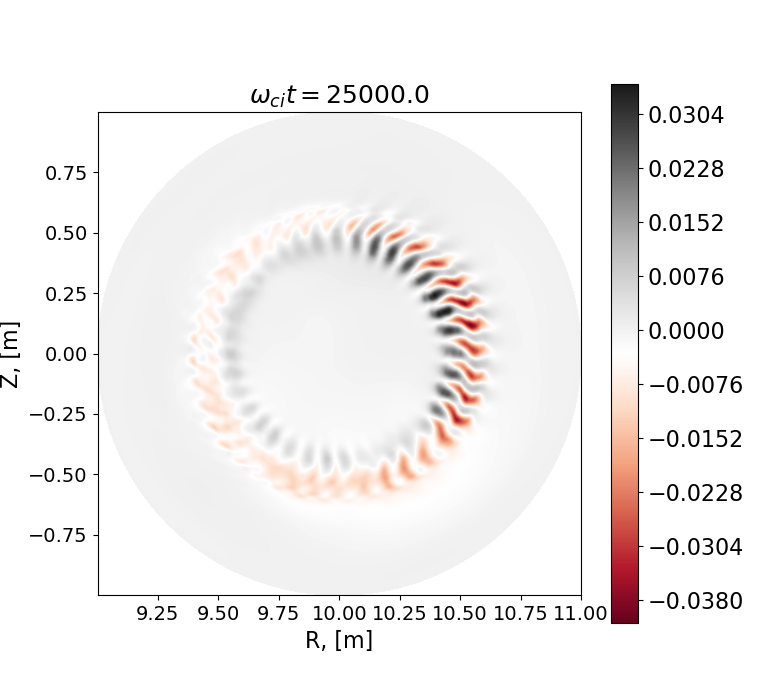} %{{./graph/BAE_b0.00025/potsc_t25e3.png}}
\includegraphics[width=0.33\textwidth]{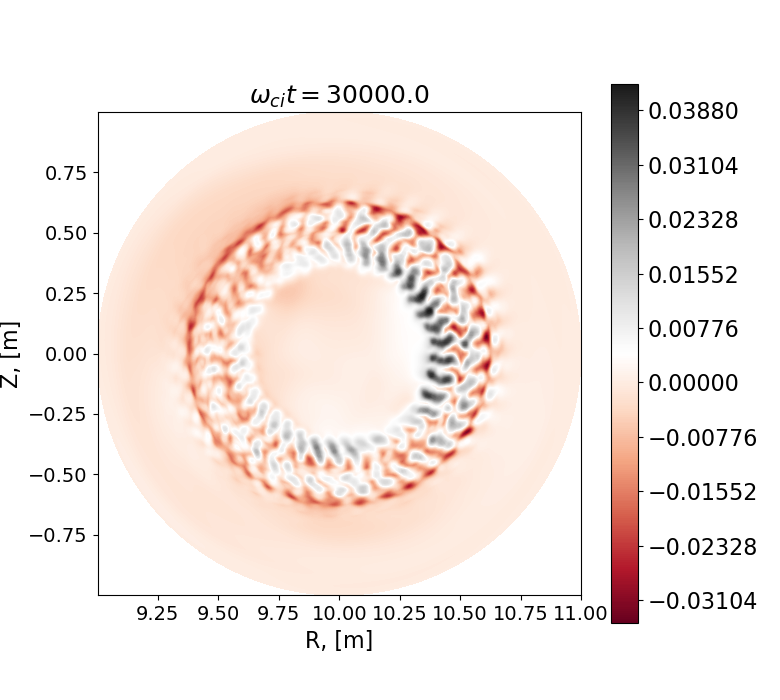} %{{./graph/BAE_b0.00025/potsc_t30e3.png}}\\
\includegraphics[width=0.33\textwidth]{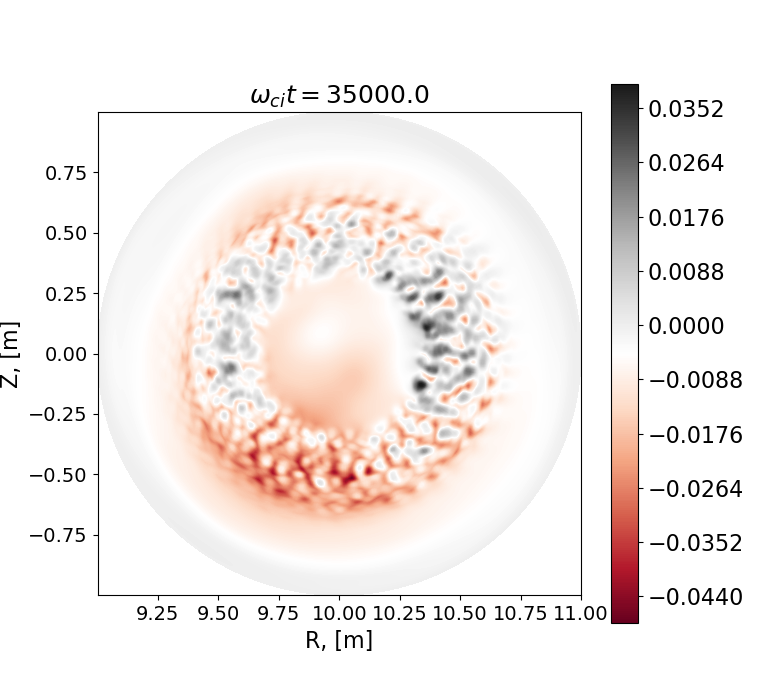} %{{./graph/BAE_b0.00025/potsc_t35e3.png}}
\includegraphics[width=0.33\textwidth]{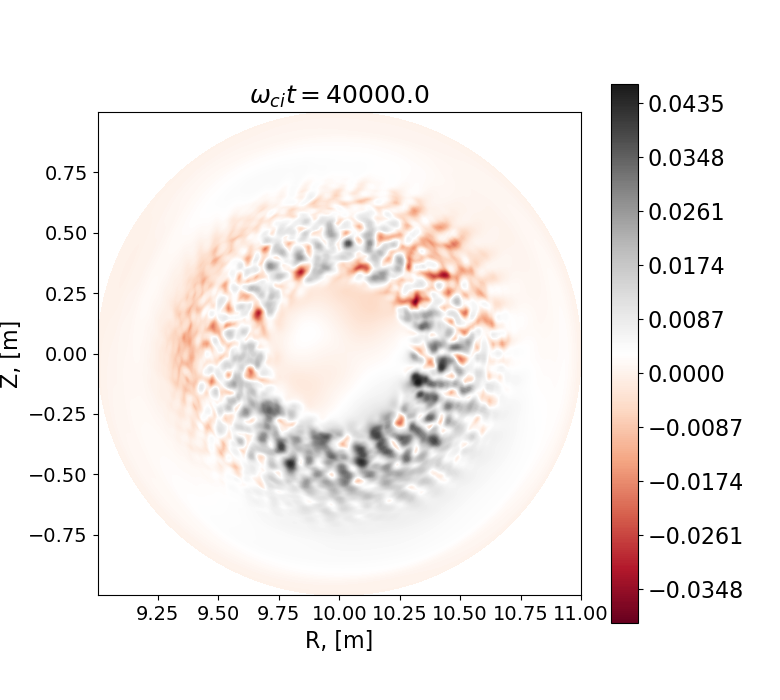} %{{./graph/BAE_b0.00025/potsc_t40e3.png}}
\includegraphics[width=0.33\textwidth]{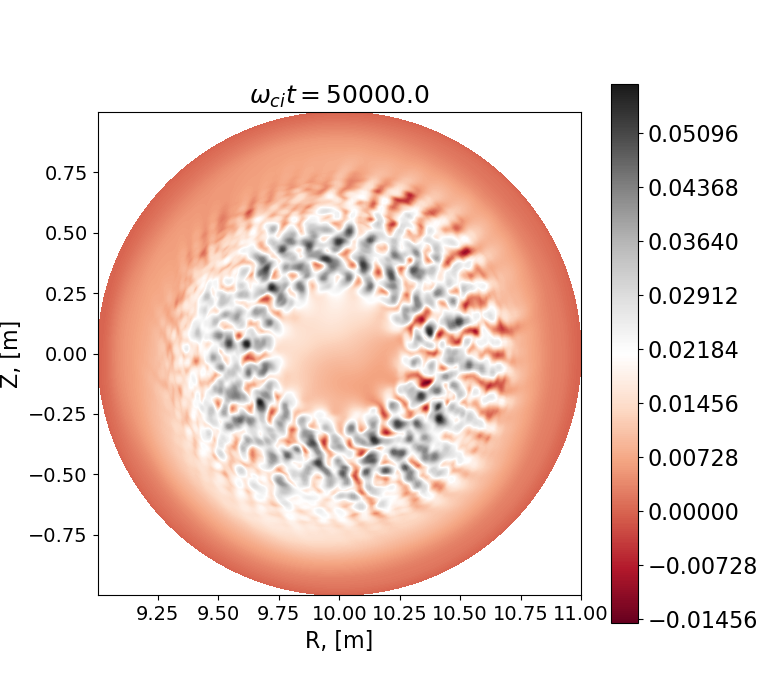} %{{./graph/BAE_b0.00025/potsc_t50e3.png}}
\caption{\Brunner{
Evolution of the electrostatic potential in the ITG regime ($\beta = 0.1\%$). One sees the characteristic pattern of the turbulent eddies decorrelated by the zonal flow.
}}
\label{potsc_ITG}
\end{figure}
\begin{figure}
\includegraphics[width=0.33\textwidth]{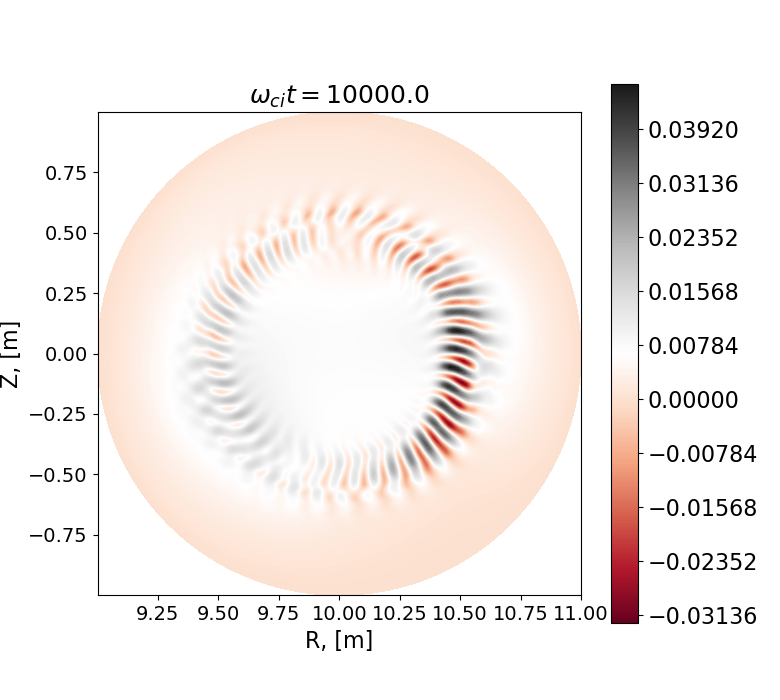} %{{./graph/BAE_b0.004/potsc_t10e3.png}}
\includegraphics[width=0.33\textwidth]{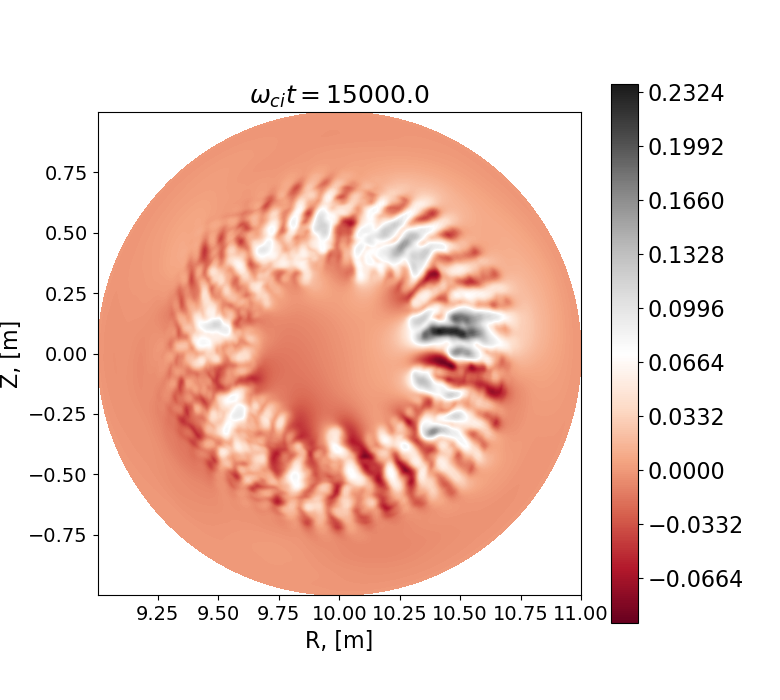} %{{./graph/BAE_b0.004/potsc_t15e3.png}}
\includegraphics[width=0.33\textwidth]{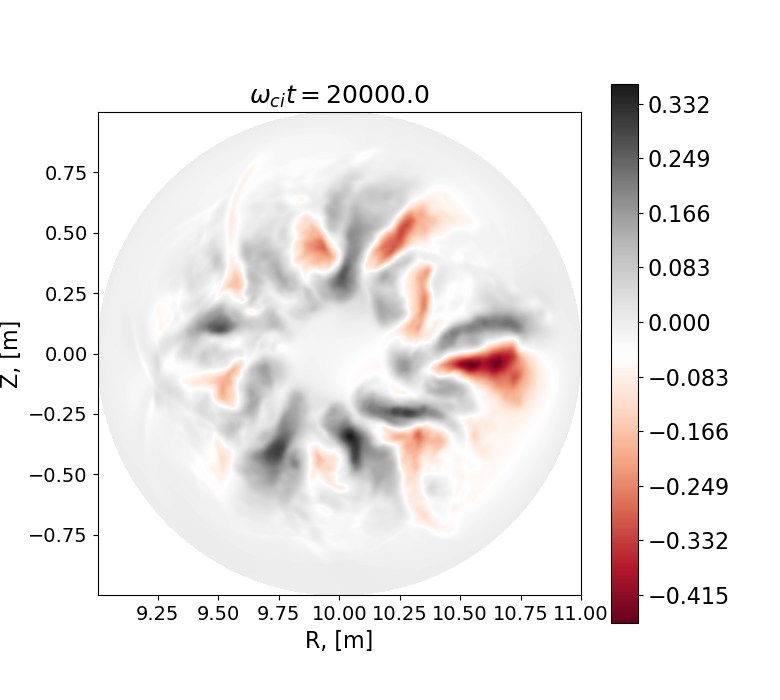} %{{./graph/BAE_b0.004/potsc_t20e3.png}} \\
\includegraphics[width=0.33\textwidth]{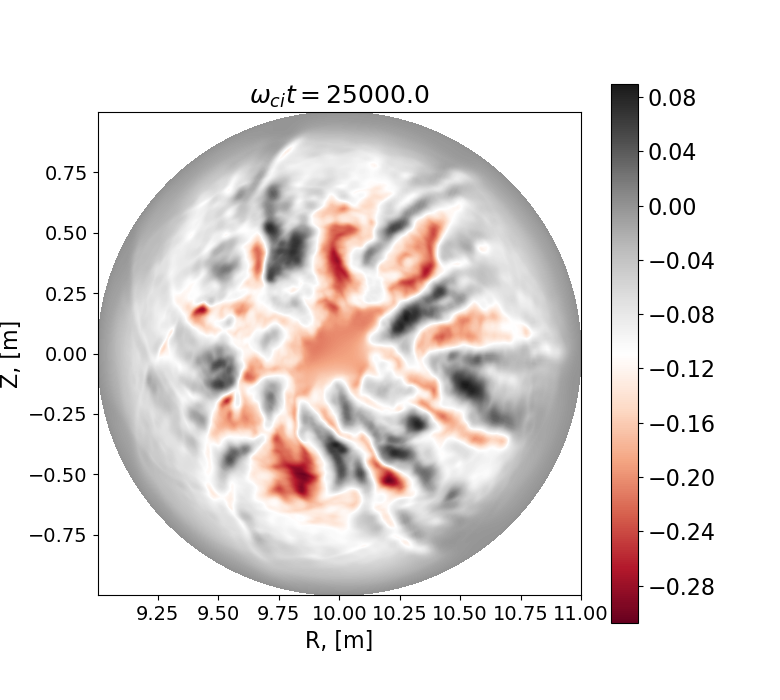} %{{./graph/BAE_b0.004/potsc_t25e3.png}}
\includegraphics[width=0.33\textwidth]{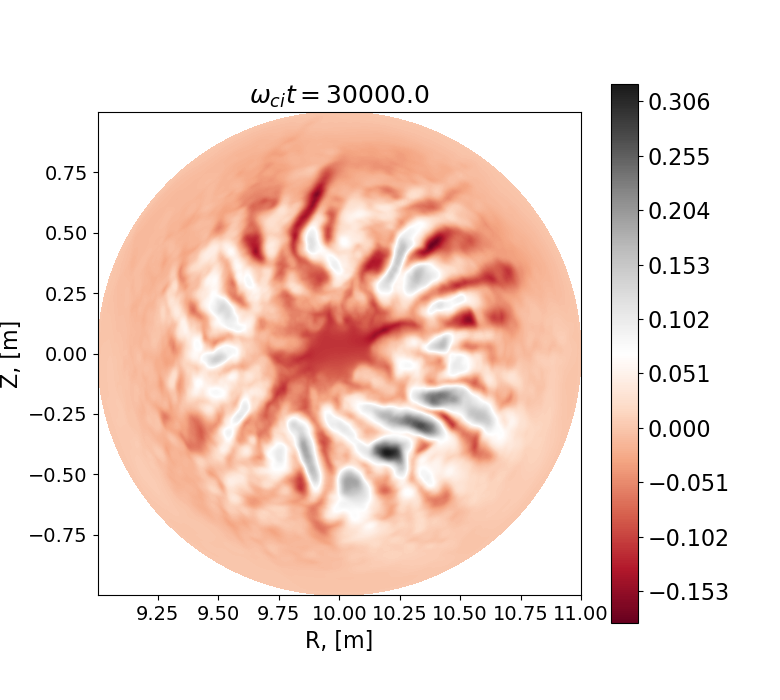} %{{./graph/BAE_b0.004/potsc_t30e3.png}}
\includegraphics[width=0.33\textwidth]{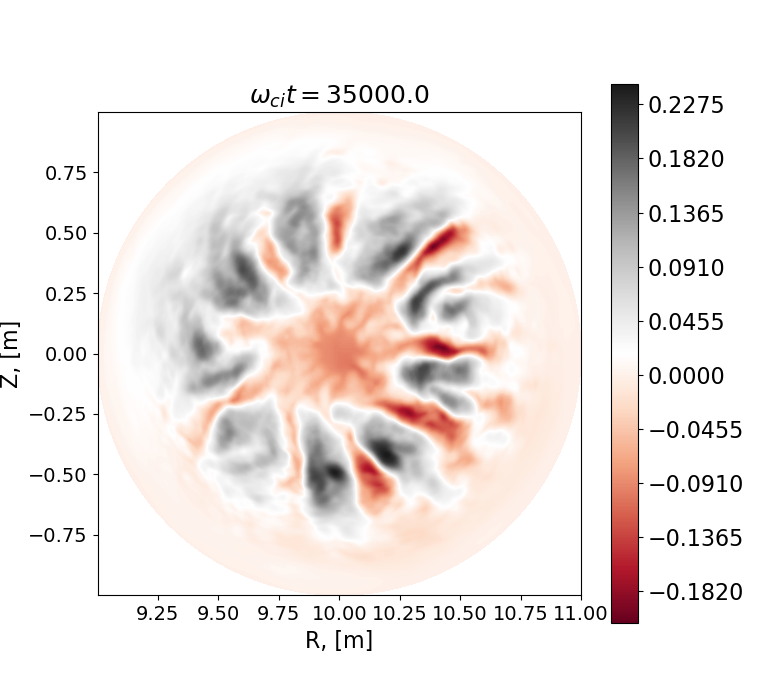} %{{./graph/BAE_b0.004/potsc_t35e3.png}}
\caption{\Brunner{
Evolution of the electrostatic potential in the KBM regime ($\beta = 1.6\%$). Global mode structure dominates the turbulence in the nonlinear phase. The poloidal periodicity of this structure corresponds to the BAE instability \cite{Biancalani_JPP} with the toroidal mode number $n = 5$ and the poloidal mode number $m = 9$.
}}
\label{potsc_KBM}
\end{figure}
\begin{figure}
\includegraphics[width=0.33\textwidth]{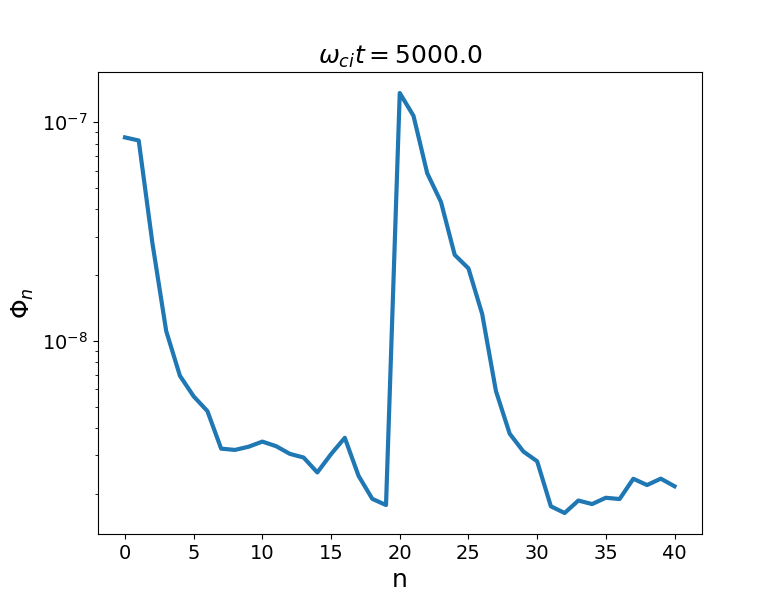} %{{./graph/BAE_b0.00025/spectr_n_t05e3.png}}
\includegraphics[width=0.33\textwidth]{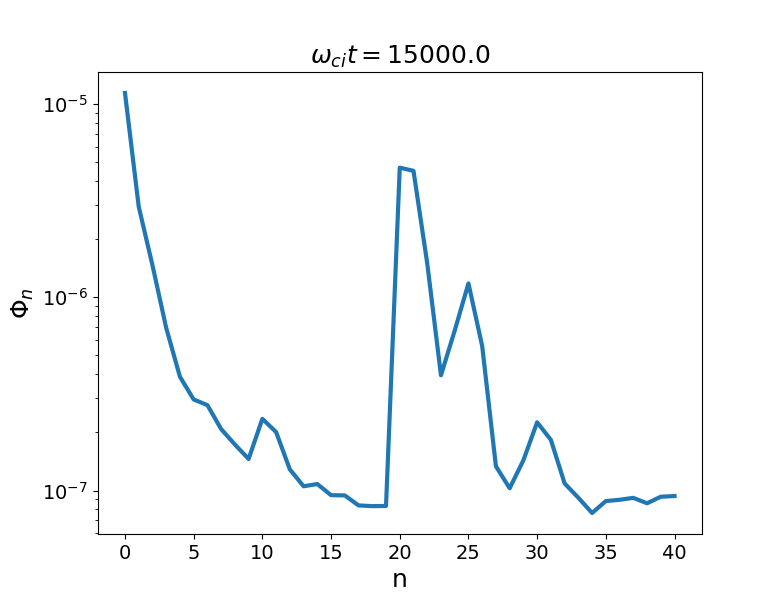} %{{./graph/BAE_b0.00025/spectr_n_t15e3.png}}
\includegraphics[width=0.33\textwidth]{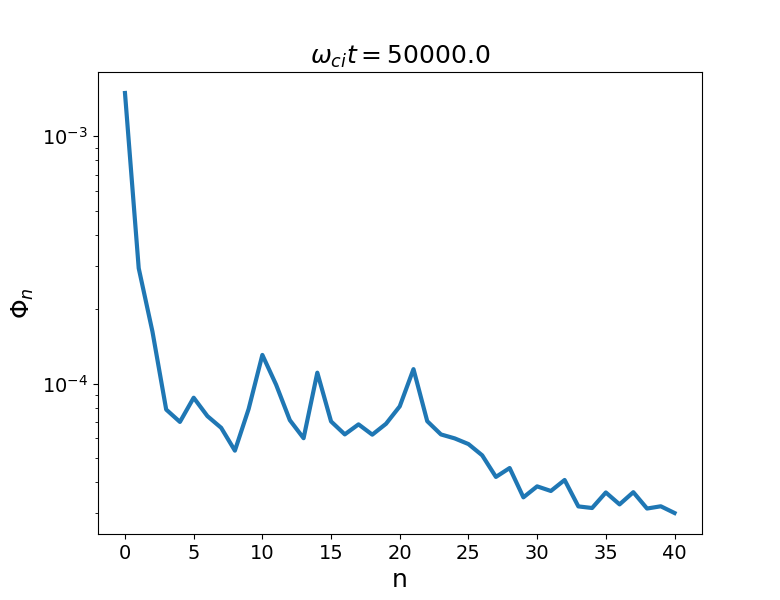} %{{./graph/BAE_b0.00025/spectr_n_t50e3.png}}
\caption{\Brunner{
Evolution of the electrostatic toroidal spectrum in the ITG regime ($\beta = 0.1\%$). The evolution starts with the spectrum peaking at the numbers $n \sim 22$. It evolves into the state with the zonal harmonic $n = 0$, ITG component $n \sim 22$, and the BAE component $n = 5$ as well as BAE's nonlinear harmonics $n = 10$ and $n = 15$.
}}
\label{spectrum_ITG}
\end{figure}
\begin{figure}
\includegraphics[width=0.33\textwidth]{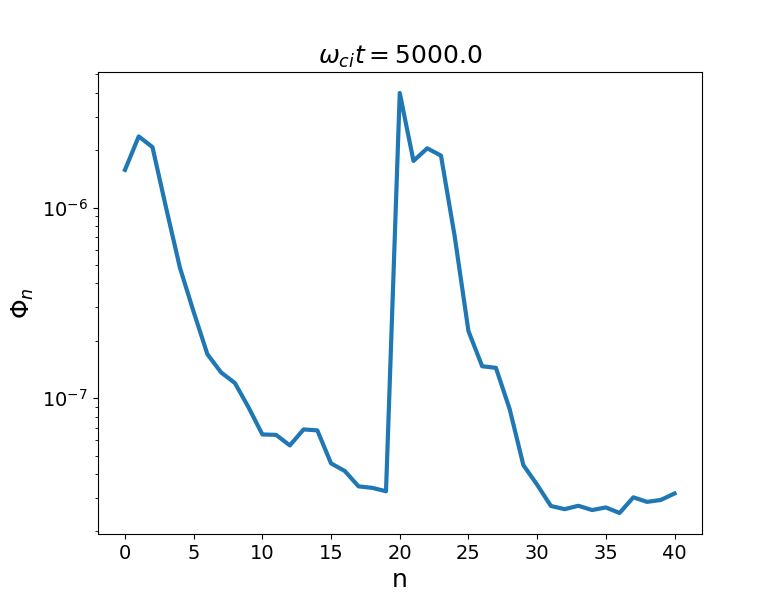} %{{./graph/BAE_b0.004/spectr_n_t05e3.png}}
\includegraphics[width=0.33\textwidth]{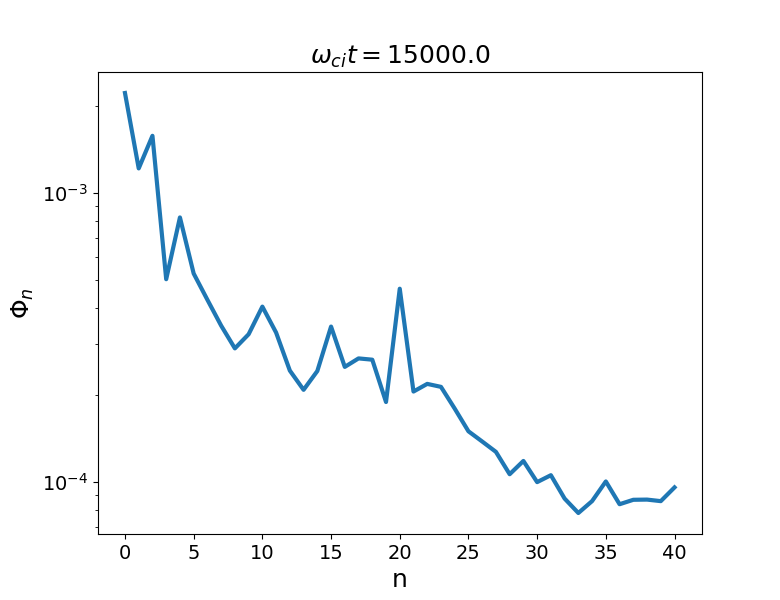} %{{./graph/BAE_b0.004/spectr_n_t15e3.png}}
\includegraphics[width=0.33\textwidth]{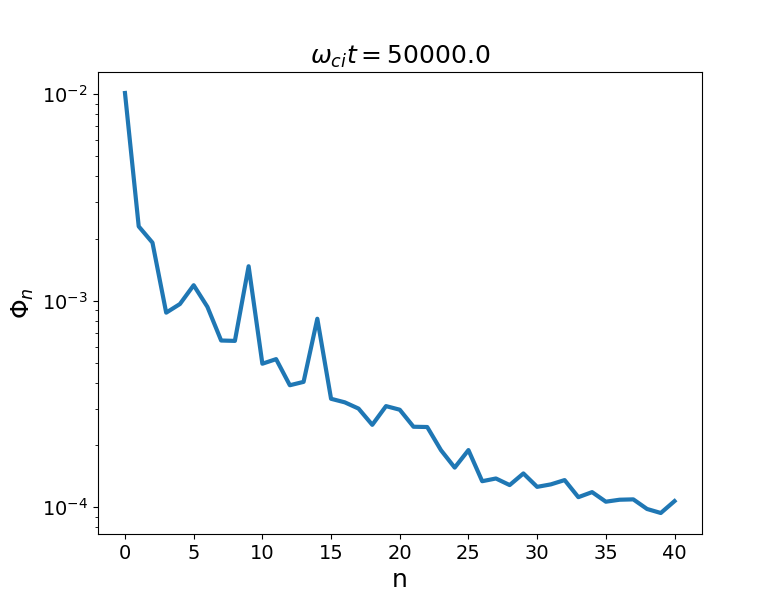} %{{./graph/BAE_b0.004/spectr_n_t50e3.png}}
\caption{\Brunner{
Evolution of the electrostatic toroidal spectrum in the KBM regime ($\beta = 1.6\%$). The evolution starts with the spectrum peaking at the mode numbers $n \sim 22$. It evolves into the state with the zonal harmonic $n = 0$ and quite strong BAE component $n = 5$ with its nonlinear harmonics $n = 10$ and $n = 15$.
}}
\label{spectrum_KBM}
\end{figure}

The mode structure evolution of the ITG turbulence, Fig.~\ref{potsc_ITG}, shows a typical pattern of the turbulent eddies decorrelated by the zonal flow. In contrast, the nonlinear phase of the high-beta case is dominated by a global mode. The poloidal periodicity of this structure corresponds to the beta-induced Alfv\'en Eigenmode (BAE) instability considered in Ref.~\cite{Biancalani_JPP} which has the toroidal mode number $n = 5$ and the poloidal mode number $m = 9$. The fast ions have been included in Ref.~\cite{Biancalani_JPP} but it can be shown that this BAE mode is linearly unstable even in the absence of the fast particles destabilized by the background plasma gradients with the growth rate increasing with the plasma beta.
\Sanchez{At the lower beta, the global structure with the BAE periodicity appears when the fast particles (driving the BAE) are included \cite{Biancalani}.}
The BAE spectral component is clearly observed in the high-beta spectrum shown in Fig.~\ref{spectrum_KBM}. Here, the absolute value of the electrostatic potential corresponding to a given toroidal mode number and averaged over all radial positions and poloidal mode numbers is shown. One sees how the spectrum, initiated at the mode numbers $20 < n < 30$ evolves into the spectrum dominated by $n = 0$ and the toroidal mode numbers corresponding to the BAE, the principal mode number $n = 5$ and its nonlinear harmonics $n = 10$ and $n = 15$. Interestingly, the low-beta spectrum, shown in Fig.~\ref{spectrum_ITG}, also evolves into the state which clearly contains the BAE component although the mode numbers characteristic to the ITG turbulence, $20 < n < 30$ can also be clearly seen. This is in contrast to the high-beta case Fig.~\ref{spectrum_KBM} where this part of the spectrum is suppressed.

\begin{figure}
\includegraphics[width=0.47\textwidth]{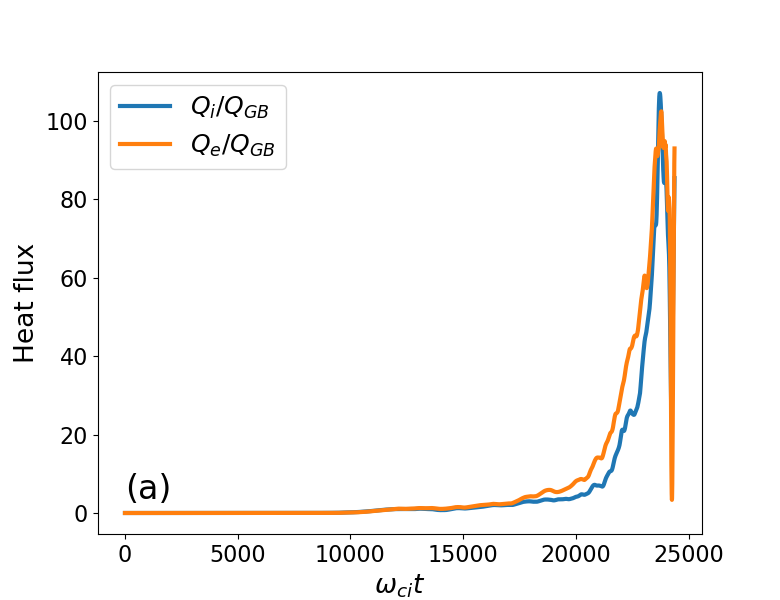} %{{./graph/hflux_t_pblin.png}}
\includegraphics[width=0.47\textwidth]{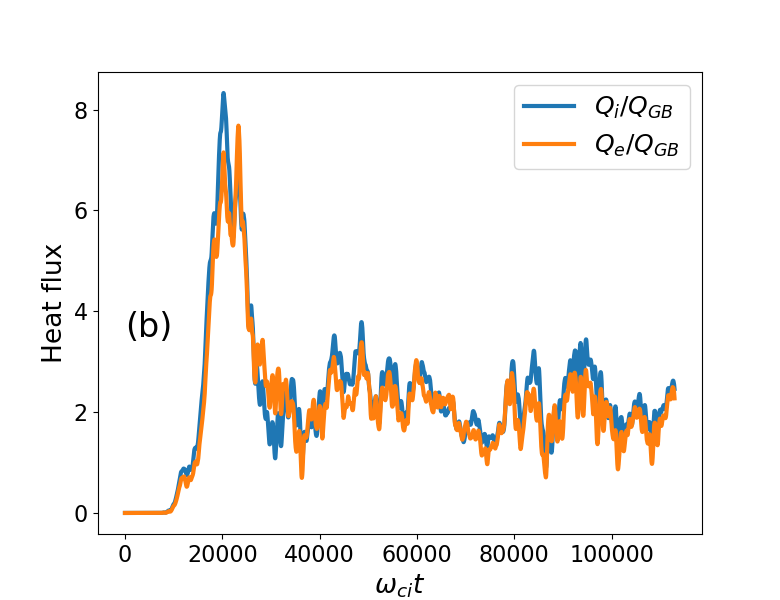} %{{./graph/hflux_t_pblin_bpert.png}}
\caption{\Sanchez{
(a) The averaged radial heat flux density for $\beta = 1.6\%$ with the linearized pullback, Eq.~(\ref{ic_f_lin}), and the magnetic-flutter nonlinearity neglected $\vc{b}^* = \vc{b}^*_0$. This simulation crashes. (b) The same paramters as in (a) with the only difference that the magnetic-flutter nonlinearity is included meaning that all terms in Eq.~(\ref{bstar}) are included in $\vc{b}^*$ definition. This simulation runs without any numerical instability.
}}
\label{hflux_t_pblin}
\end{figure}

\Sanchez{
We will conclude this section with some observation concerning the numerical peculiarities of the high-beta electromagnetic simulations. First, we address the role of the "nonlinear pullback" in the turbulence saturation.
In Fig.~\ref{hflux_t_pblin}(a), it can be seen how a high-beta simulation crashes in the nonlinear phase if the linearized pullback approximation, Eq.~(\ref{ic_f_lin}), and the approximation $\vc{b}^* =  \vc{b}^*_0$ in Eq.~(\ref{bstar}) are used. In contrast, these approximations have a marginal effect in the low-beta case (electromagnetic ITG regime) considered in Ref.~\cite{Biancalani}. It is important to understand which of these two approximations is responsible for the crash seen in Fig.~\ref{hflux_t_pblin}(a). For this purpose, we run a simulation which employs the linearized transformation, Eq.~(\ref{ic_f_lin}) but uses all terms in Eq.~(\ref{bstar}), including the flutter perturbation (the term proportional to $\nabla\Bgav{\As}$). In Fig.~\ref{hflux_t_pblin}(b), we observe that this simulation runs without any numerical instability. The linearized pullback transformation, Eq.~(\ref{ic_f_lin}) applied in Fig.~\ref{hflux_t_pblin}(b), does not seem to make any substantial difference. This indicates importance of the nonlinear magnetic flutter for the Alfv\'enic turbulence saturation at higher beta.
}

\begin{figure}
\includegraphics[width=0.47\textwidth]{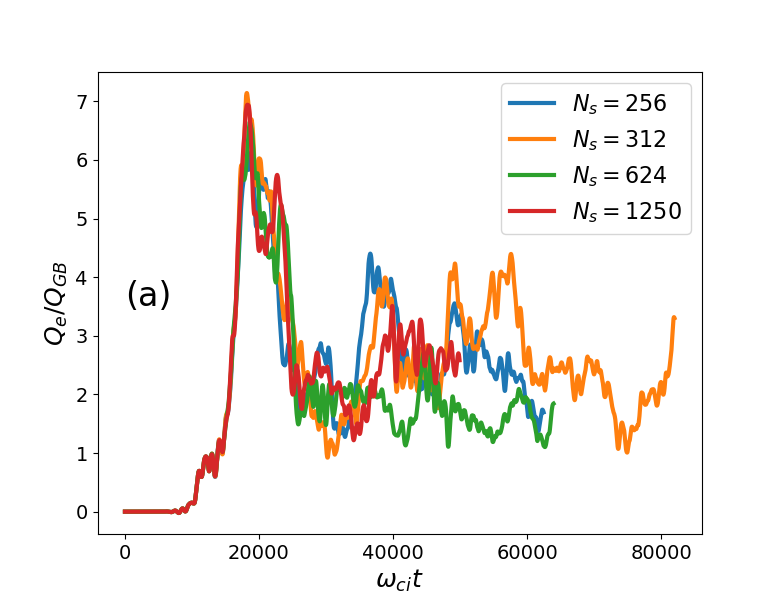} %{{./graph/hflux_comp_t-b0.004.png}}
\includegraphics[width=0.47\textwidth]{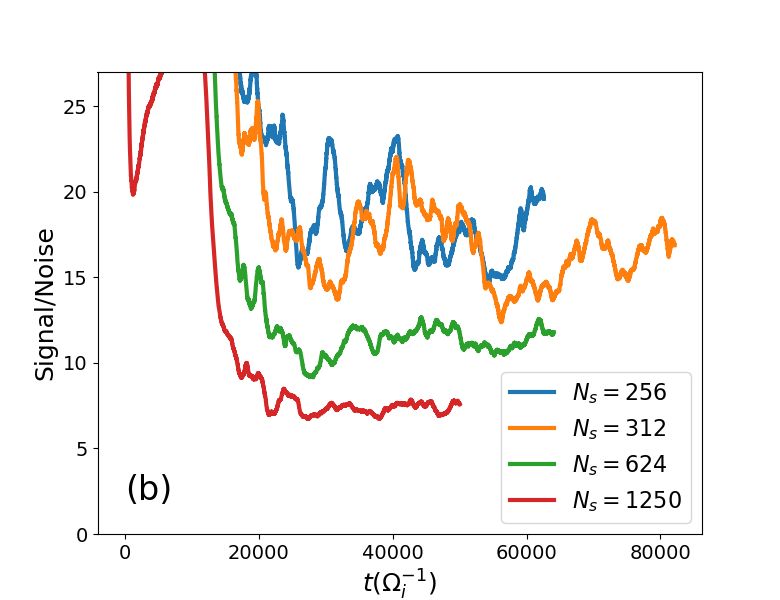} %{{./graph/signal_comp_t-b0.004-fsig.png}}
\caption{(a) The heat-flux density for $\beta = 1.6\%$, $\gamma_{Krook}/\omega_{ci} = 10^{-4}$, and different radial resolutions. (b) The signal-to-noise ratio (same definition as in Ref.~\cite{Sanchez_EUTERPE}).}
\label{eflux_b0.004_ns}
\end{figure}

The second numerical point to address here is related to the radial resolution necessary for the electromagnetic turbulence simulations. It was actively discussed in the early phase of the numerical electromagnetic gyrokinetics that resolution of the electron skin depth is needed for such simulations to be numerically stable \cite{YChen_skindepth,Jenko_skindepth}. Such a requirement would make electromagnetic simulations very difficult, in particular for large machines such as ITER, where the radial resolution $N_s \approx 40\,000$ would be needed. Fortunately, in our high-beta simulations we see that resolving the collisionless electron skin depth is not needed for the numerical stability. All the simulations shown so far had the radial resolution $N_s = 256$ which is significantly below the electron skin depth resolution. For example, to resolve the skin depth for $\beta = 1.6\%$, $N_s = 1250$ would be required. In Fig.~\ref{eflux_b0.004_ns}(a), the heat flux is plotted for for $\beta = 1.6\%$ and different radial resolutions, including the case with the electron skin depth being resolved. One sees that the heat fluxes are very close during the overshoot and comparable later for all radial resolutions considered.
\Sanchez{In these simulations, the total number of markers did not change although it would have to be increased when using the larger grid resolutions.
The problem was in the GPU memory limitations prohibiting increasing the marker resolution for the larger radial grids. The expected consequence is the increase in the "noise" level shown in Fig.~\ref{eflux_b0.004_ns}(b). The signal corresponds here to the electrostatic potential inside the diagonal Fourier filter \cite{Jolliet_ORB5} with the half-width $\Delta m$  included in the simulations (e.~g.~to push the markers). The noise corresponds to the electrostatic potential outside this filter. Normally, $\Delta m = 5$ except in Fig.~\ref{eflux_b0.004_ns}, where $\Delta m = 3$.}

\Sanchez{
The noise generation is the third numerical aspect to mention here. In Fig.~\ref{eflux_b0.004_ns}(b), we observe that the noise level is not negligible although one cannot say that the simulation is dominated by the noise. For the typical parameters used here, the radial grid resolution $N_s = 256$, the number of the ion markers $N_i = 200$ million, and the number of the electron markers $N_e = 500$ million, the noise level is about $5\%$ in the final phase of the simulation (it is much smaller at the beginning of the nonlinear phase). The noise level can be reduced by increasing the number of the markers or increasing the decay rate of the Krook operator (affecting the linear growth rate). We can reduce the noise level even for $\gamma_{Krook} = \gamma_{lin}/10$ down to about $4\%$ by increasing the total number of the markers to $N_i = 400$ million and $N_e = 1.2$ billion. To run this case on Marconi100 GPUs, we needed $64$ nodes. %which is a substantial allocation for this computer system.
%The technical problem here is again related to the GPU memory: $N_i = 100\,000\,000$ fits in $16$ nodes for a moderate grid resolution (increasing the memory usage results in a CUDA crash).
We shall note that this paper represents the first attempt to simulate electromagnetic turbulence in the high-beta regime using ORB5. Previous works \cite{Bottino_noise,McMillan2008,Sanchez_EUTERPE} dedicated to the noise control in ORB5 simulations used the adiabatic-electron electrostatic approximation. It is clear that more effort is needed to extend these earlier results to the electromagnetic regime. These effort shall address alternative noise-control schemes (such as the "quadtree" approach \cite{Sanchez_EUTERPE}), conservation properties of the noise-control techniques in the electromagnetic regime, effects of the energy and particle sources, as well as dependence of the noise level on the physical parameters such as the plasma beta, the mass ratio, the machine size $L_x$, and the magnetic equilibrium. This program goes beyond the scope of this present paper.
}
%
%%%%%%%%%%%%%%%%%%%%%%%%%%%%%%%%%%%%%%%%%%%%%%%%%%%%%%%%%
%
\subsection{Increasing machine size and mass ratio}
Another limitation of Ref.~\cite{Biancalani} was the rather small machine size measured in gyroradii $L_x = 2 r_a/\rho_i = 350$ (for ITER $L_x > 1000$). Increasing $L_x$ is computationally more challenging than increasing $\beta$ since it requires more resolution. We typically include about $L_x/3$ poloidal modes in our simulations, covering roughly the range $0 \le k_{\perp}\rho_i \le 1$. The number of the toroidal modes are defined by the condition of small $k_{\|}$ for a given poloidal mode number (i.~e.~small Landau damping) which implies $m_{\max} \approx q_{\max} n_{\max}$ where $q_{\max}$ is the maximum value of the safety factor. These broad turbulence spectra require a spatial grid fine enough to resolve all the modes included (typically four grid points per Fourier mode both in the poloidal and in the toroidal directions). Finally, the radial resolution and the number of the markers for all species need to be adjusted (increased) accordingly to fulfill the condition $0 \le k_{\perp}\rho_i \le 1$ and to have enough markers per Fourier mode and per radial grid cell. These resolution requirements may be in conflict with the available resources, such as the GPU memory, limiting both the number of the markers and the number of the grid points.

\begin{figure}
\includegraphics[width=0.47\textwidth]{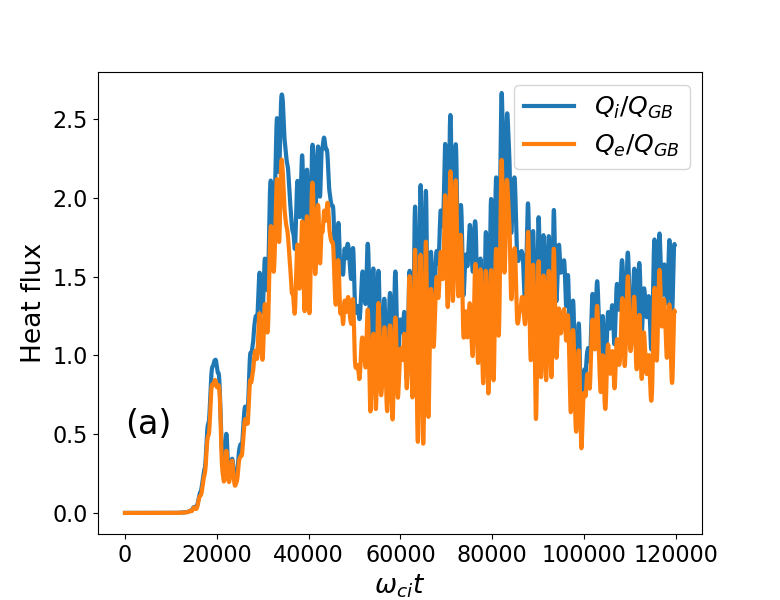} %{{./graph/BAE_b0.002_lx480_np4e8/hflux_t_np4e8.png}}
\includegraphics[width=0.47\textwidth]{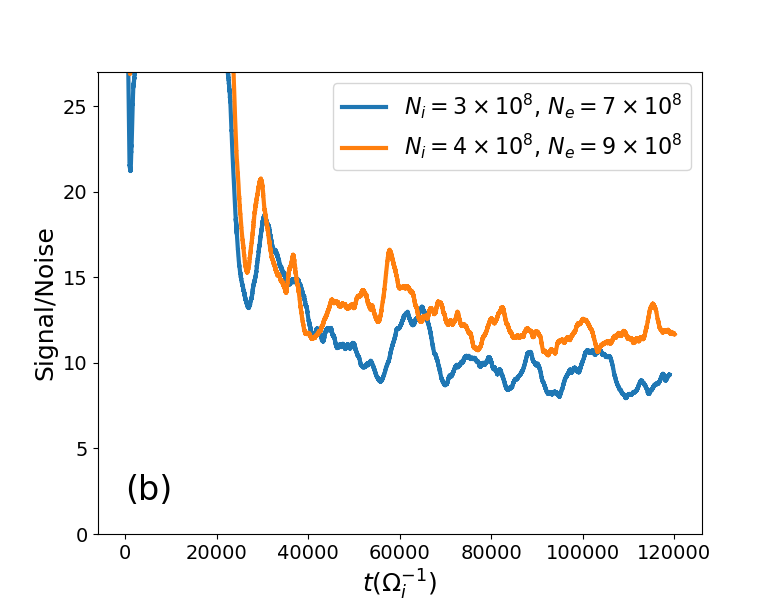} %{{./graph/BAE_b0.002_lx480_np4e8/signal_comp_t-fsig.png}}
\caption{(a) The heat flux for $\beta = 0.8\%$, $L_x = 480$, $N_i = 4 \times 10^8$, and $N_e = 9 \times 10^8$. (b) The signal-to-noise ratio for $\beta = 0.8\%$, $L_x = 480$, and different marker resolutions.}
\label{b0.004_lx480}
\end{figure}
\begin{figure}
\includegraphics[width=0.33\textwidth]{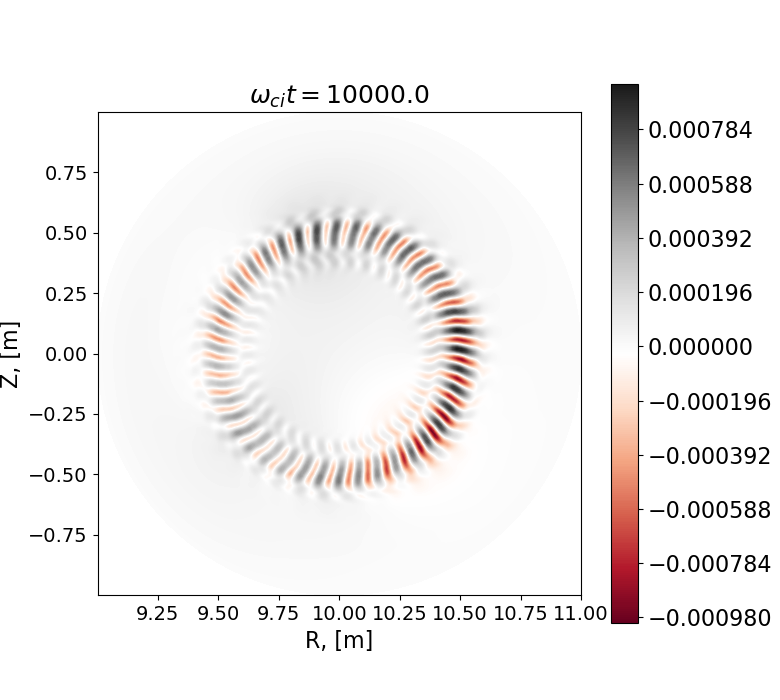} %{{./graph/BAE_b0.002_lx480_np4e8/potsc_t10e3.png}}
\includegraphics[width=0.33\textwidth]{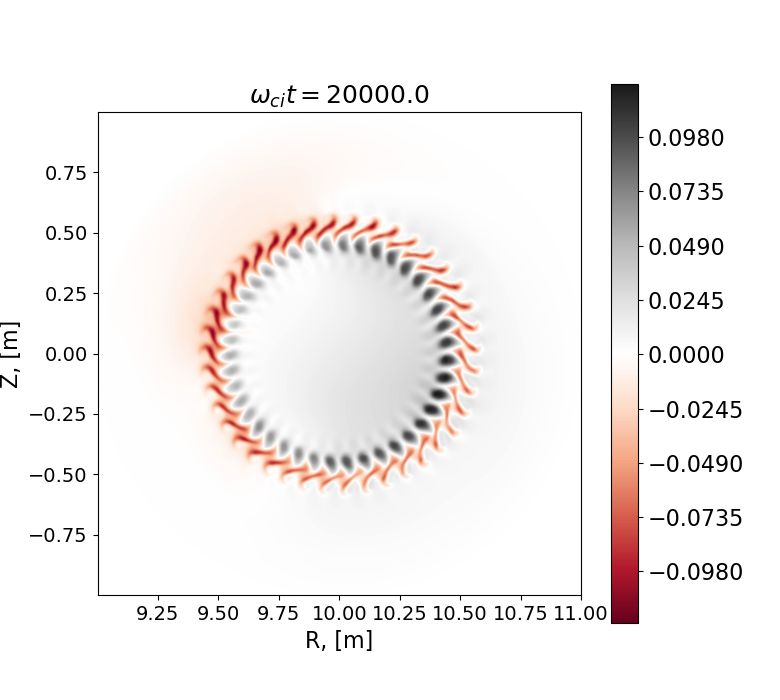} %{{./graph/BAE_b0.002_lx480_np4e8/potsc_t20e3.png}}
\includegraphics[width=0.33\textwidth]{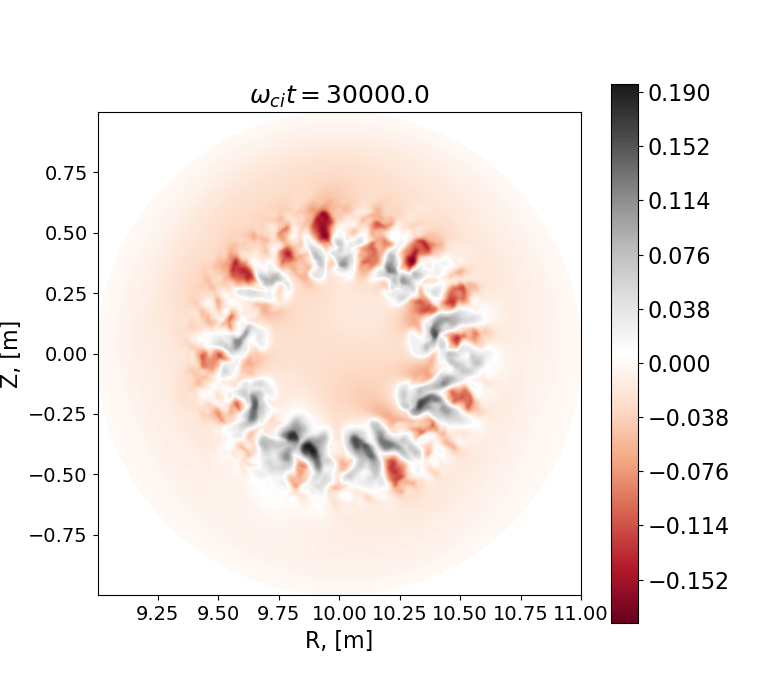} %{{./graph/BAE_b0.002_lx480_np4e8/potsc_t30e3.png}} \\
\includegraphics[width=0.33\textwidth]{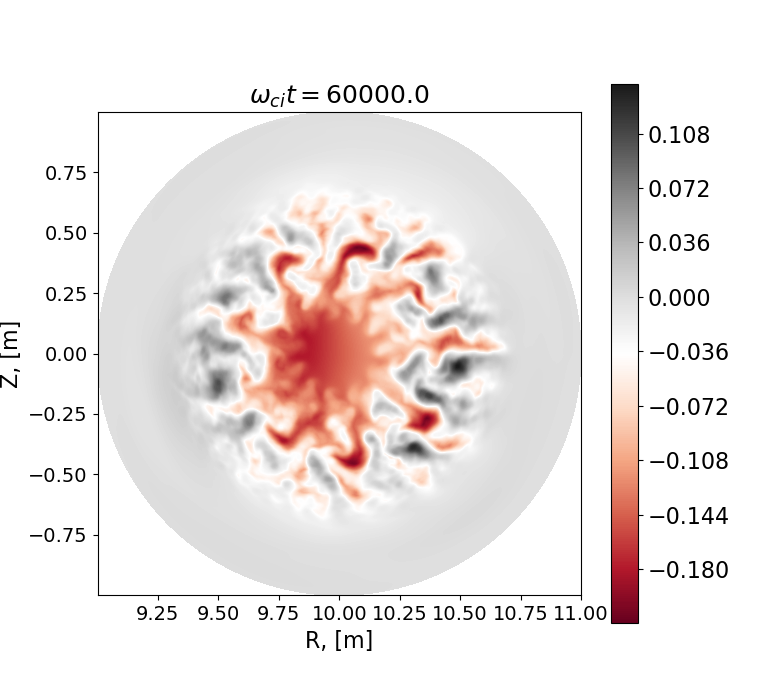} %{{./graph/BAE_b0.002_lx480_np4e8/potsc_t60e3.png}}
\includegraphics[width=0.33\textwidth]{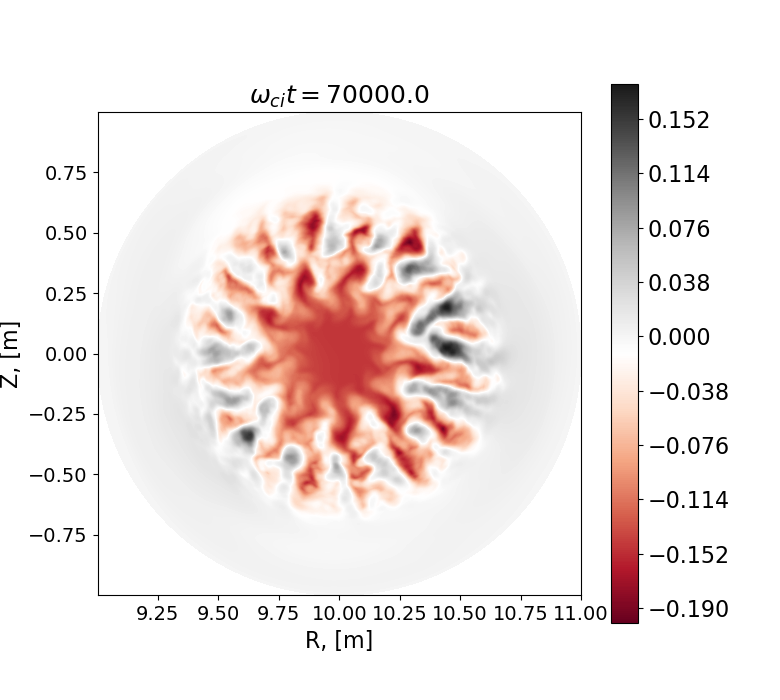} %{{./graph/BAE_b0.002_lx480_np4e8/potsc_t70e3.png}}
\includegraphics[width=0.33\textwidth]{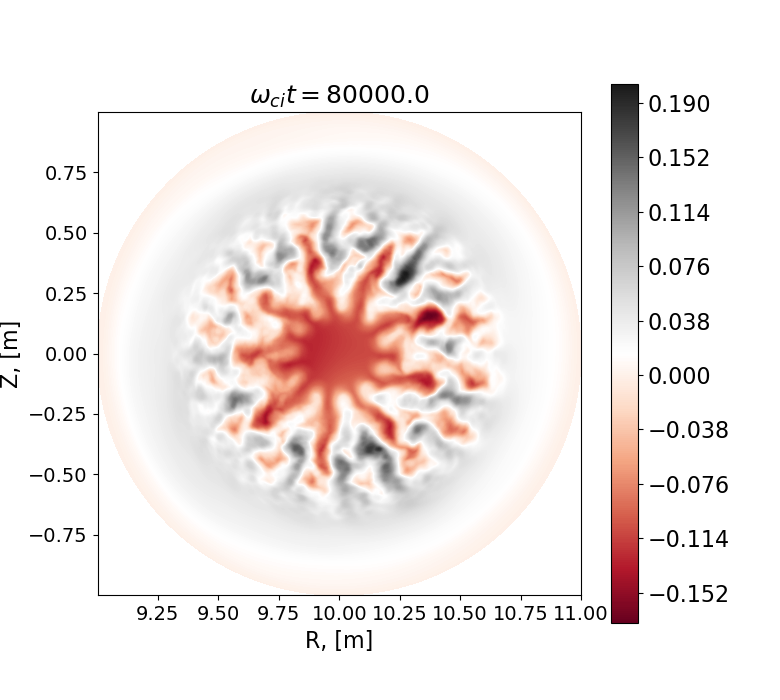} %{{./graph/BAE_b0.002_lx480_np4e8/potsc_t80e3.png}}
\caption{\Brunner{
Evolution of the electrostatic potential potential for $\beta = 0.8\%$, $L_x = 480$, $m_i/m_e = 200$, the total number of the markers $N_i = 4\times10^8$ (ions), and $N_e = 9\times10^8$ (electrons). A global mode structure (BAE) dominates the late phase of the simulation, similar to Fig.~\ref{potsc_KBM}.
}}
\label{potsc_lx480}
\end{figure}

In Fig.~\ref{b0.004_lx480}, the heat flux and the signal-to-noise ratio are shown for $\beta = 0.8\%$ and $L_x = 480$. One sees that the noise level is increased in the case with larger $L_x$. This can be expected since the number of the Fourier modes to be resolved is also larger in this case: $832$ Fourier modes for $L_x = 480$ compared to $400$ Fourier modes for $L_x = 350$ considered in the previous section. We manage to reach the noise level of $7\%$ in the late nonlinear stage in our largest simulation for $L_x = 480$ using $64$ nodes on Marconi100. The resulting electrostatic potential evolution is shown in Fig.~\ref{potsc_lx480}. Here, one can see again the characteristic global structure with the BAE periodicity emerging from the turbulent background and similar to the structure observed in Fig.~\ref{potsc_KBM}.

% \begin{figure}
% %\includegraphics[width=0.47\textwidth]{{./graph/eflux_t_b0.004_lx1050.png}}
% %\includegraphics[width=0.47\textwidth]{{./graph/potsc_surf_b0.004_lx1050.png}}
% \includegraphics[width=0.47\textwidth]{{./graph/BAE_b0.002_lx480_np4e8/temp_relaxed.png}}
% \includegraphics[width=0.47\textwidth]{{./graph/BAE_b0.002_lx480_np4e8/dens_relaxed.png}}
% \caption{(a) The heat flux and (b) the structure of the saturated electrostatic potential in the poloidal cross-section for $\beta = 1.6\%$, $L_x = 1050$, and profiles of Ref.~\cite{Biancalani}.}
% \label{b0.004_lx1050}
% \end{figure}
% %
% \begin{figure}
% %\includegraphics[width=0.47\textwidth]{{./graph/eflux_t_b0.004_lx1050.png}}
% %\includegraphics[width=0.47\textwidth]{{./graph/potsc_surf_b0.004_lx1050.png}}
% \includegraphics[width=0.47\textwidth]{{./graph/BAE_b0.002_lx480_np4e8/shearing_rate.png}}
% \includegraphics[width=0.47\textwidth]{{./graph/BAE_b0.002_lx480_np4e8/eflux_st.png}}
% \caption{(a) The heat flux and (b) the structure of the saturated electrostatic potential in the poloidal cross-section for $\beta = 1.6\%$, $L_x = 1050$, and profiles of Ref.~\cite{Biancalani}.}
% \label{b0.004_lx1050}
% \end{figure}
%

\begin{figure}
\includegraphics[width=0.47\textwidth]{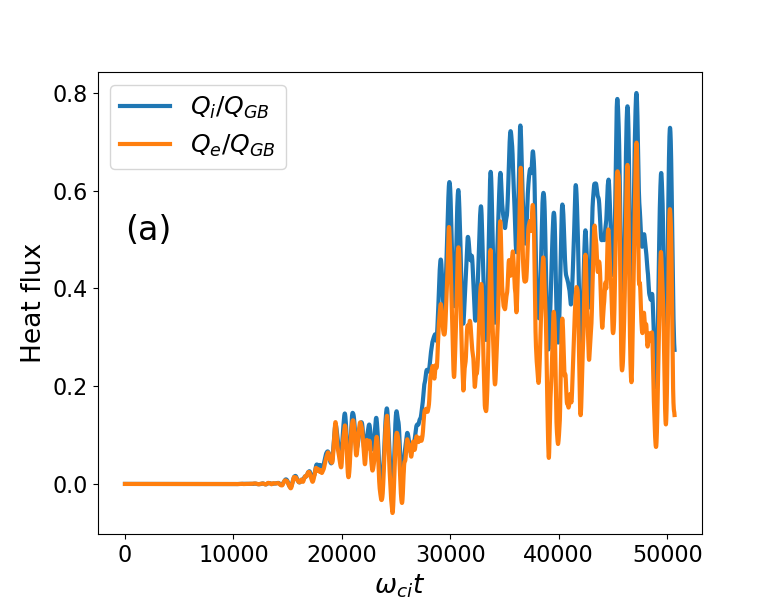} %{{./graph/BAE_b0.002_me1000/hflux_t.png}}
\includegraphics[width=0.47\textwidth]{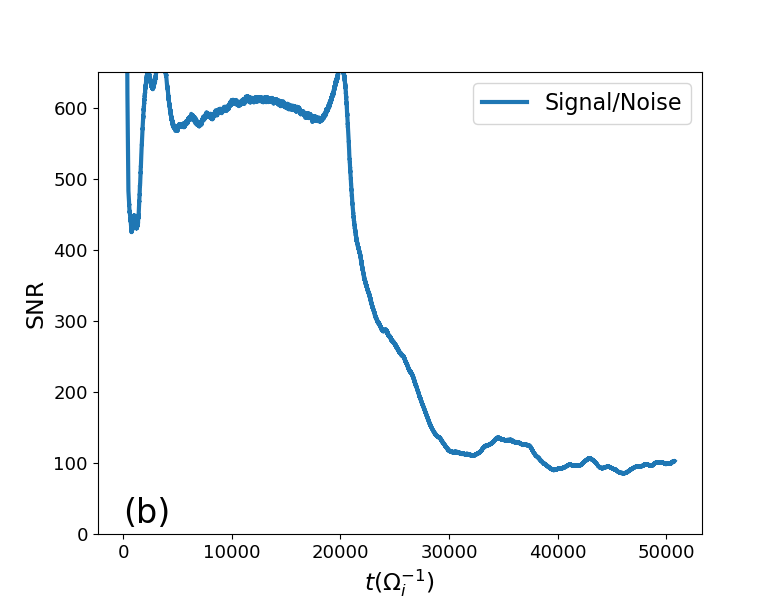} %{{./graph/BAE_b0.002_me1000/SNR_t-fsig.png}}
\caption{(a) The heat flux and (b) the signal-to-noise ratio for $\beta = 0.8\%$, $L_x = 350$, and the mass ratio $m_i/m_e = 1000$.}
\label{hflux_me1000}
\end{figure}

\begin{figure}
\includegraphics[width=0.33\textwidth]{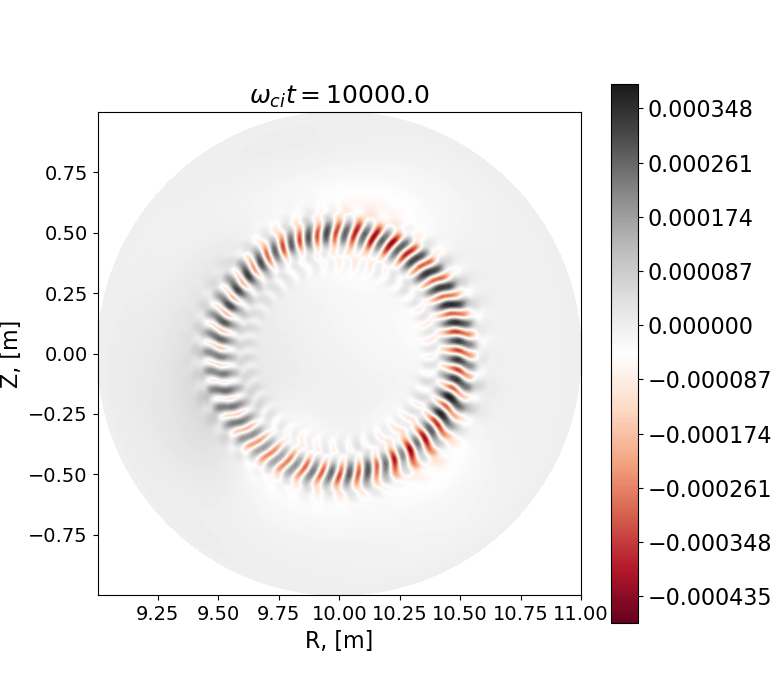} %{{./graph/BAE_b0.002_me1000/potsc_t10.png}}
\includegraphics[width=0.33\textwidth]{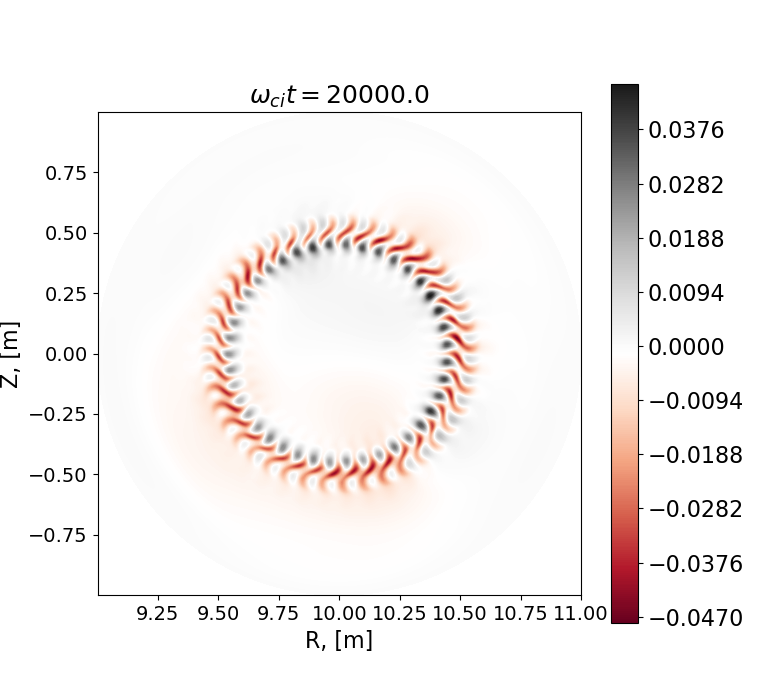} %{{./graph/BAE_b0.002_me1000/potsc_t20.png}}
\includegraphics[width=0.33\textwidth]{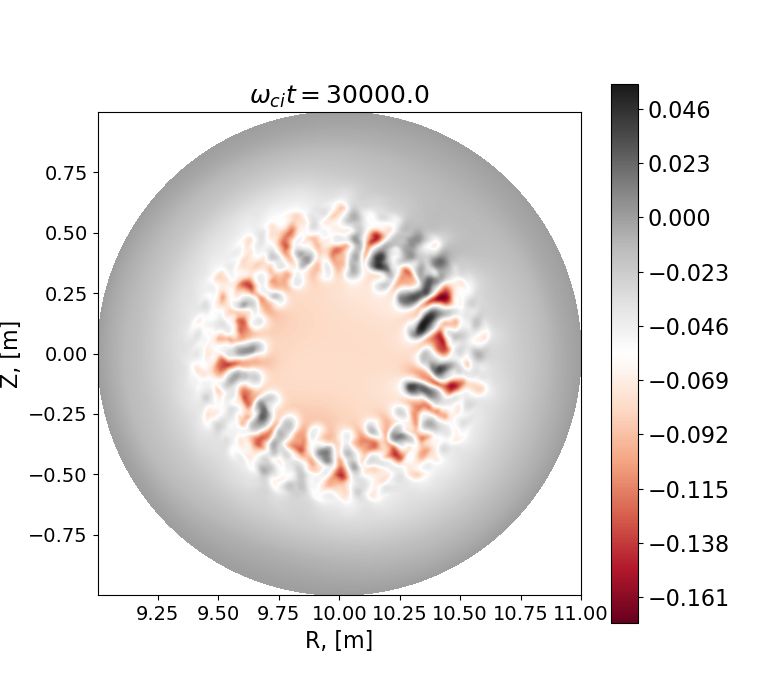} %{{./graph/BAE_b0.002_me1000/potsc_t30.png}} \\
\includegraphics[width=0.33\textwidth]{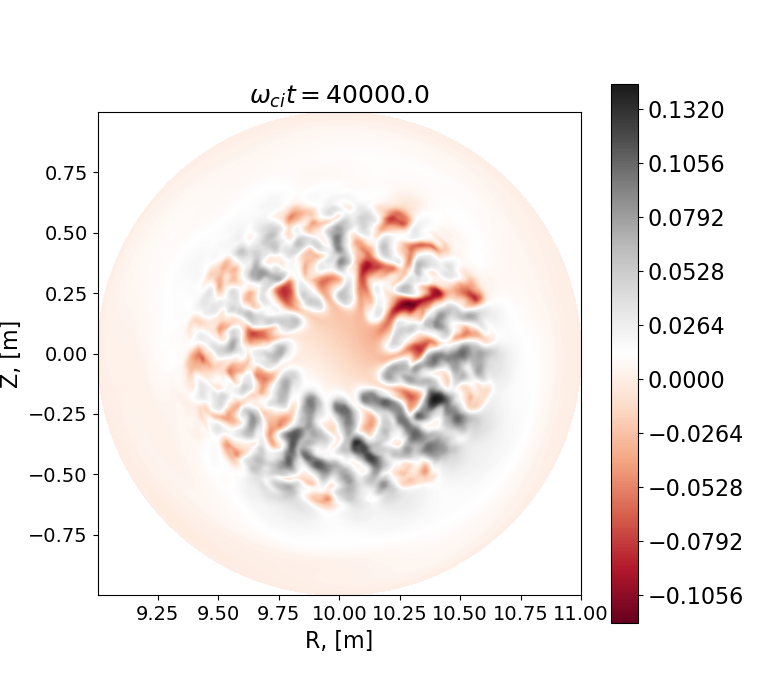} %{{./graph/BAE_b0.002_me1000/potsc_t40.png}}
\includegraphics[width=0.33\textwidth]{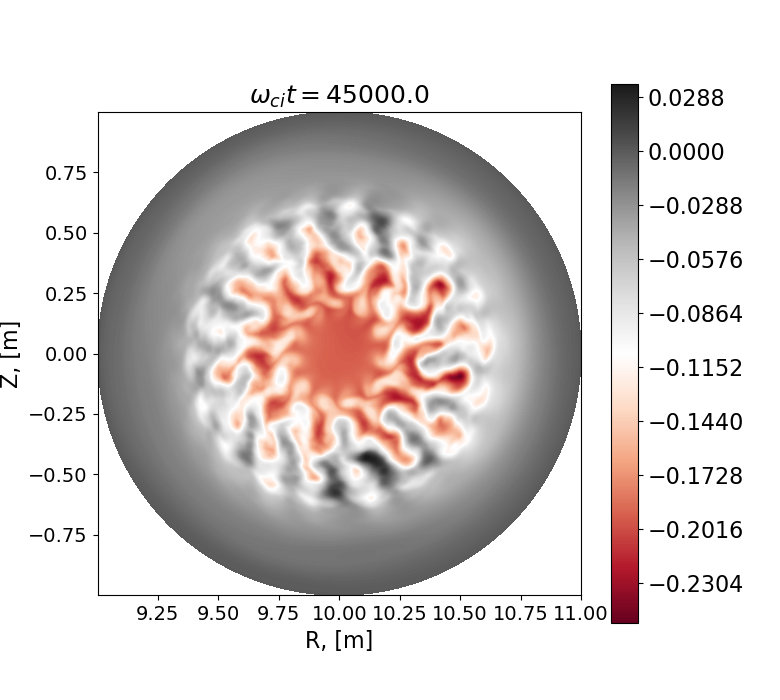} %{{./graph/BAE_b0.002_me1000/potsc_t45.png}}
\includegraphics[width=0.33\textwidth]{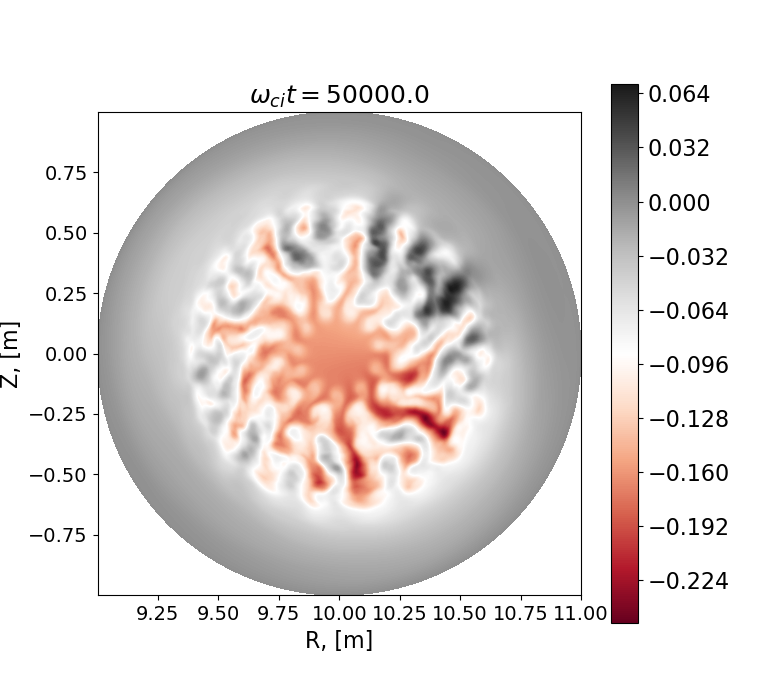} %{{./graph/BAE_b0.002_me1000/potsc_t50.png}}
\caption{\Brunner{
Evolution of the electrostatic potential for $\beta = 0.8\%$, $L_x = 350$, and the mass ratio $m_i/m_e = 1000$. Global BAE structure can be clearly seen.
}}
\label{potsc_me1000}
\end{figure}

\begin{figure}
\includegraphics[width=0.33\textwidth]{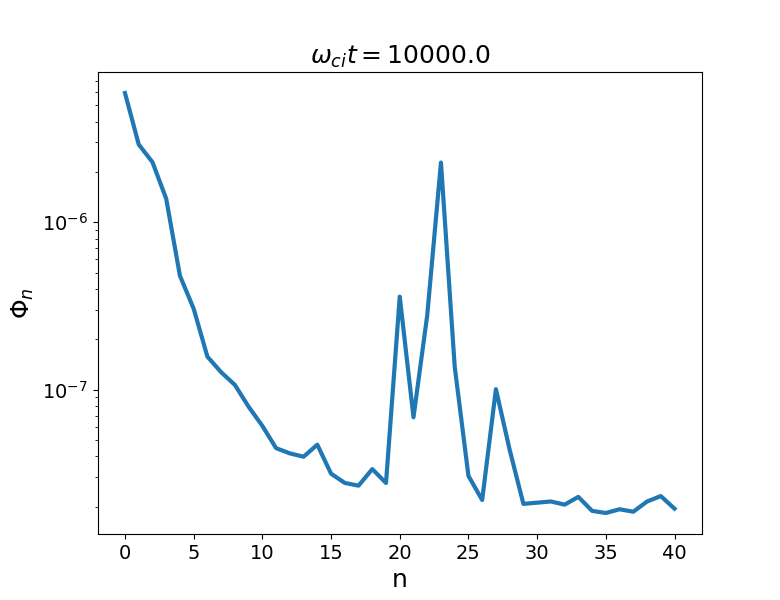} %{{./graph/BAE_b0.002_me1000/spectr_n_t10e3.png}}
\includegraphics[width=0.33\textwidth]{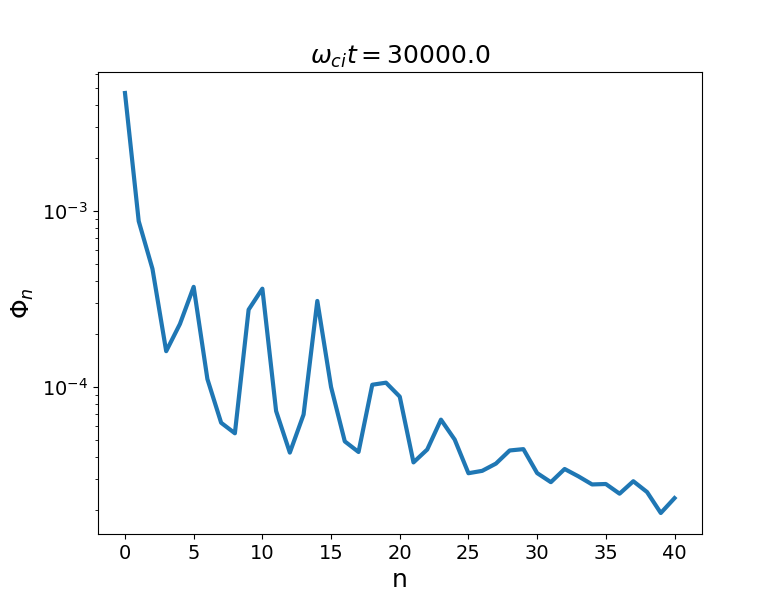} %{{./graph/BAE_b0.002_me1000/spectr_n_t30e3.png}}
\includegraphics[width=0.33\textwidth]{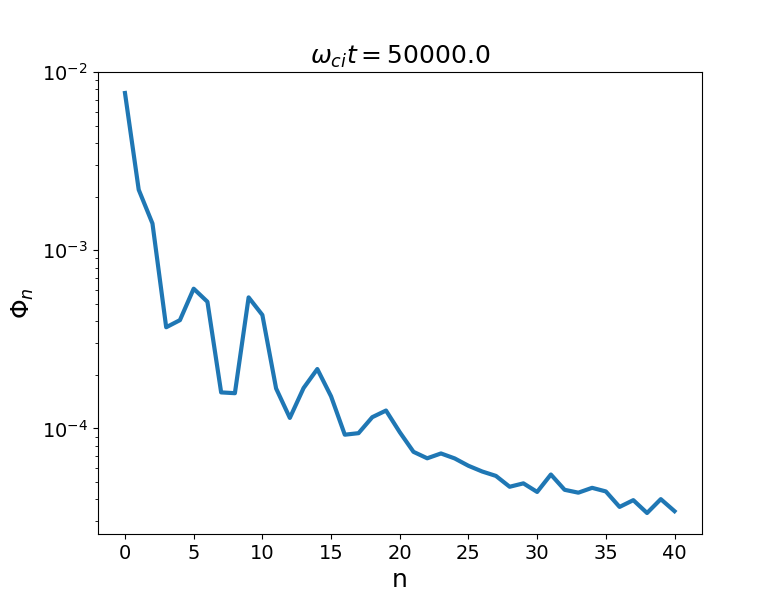} %{{./graph/BAE_b0.002_me1000/spectr_n_t50e3.png}}
\caption{\Brunner{
Evolution of the toroidal spectrum of the electrostatic potential for $\beta = 0.8\%$, $L_x = 350$, and the mass ratio $m_i/m_e = 1000$. One sees the peaks in the spectrum developing at the toroidal mode numbers corresponding to the BAE $n = 5$ and its nonlinear harmonics $n = (10, 15)$.
}}
\label{spectrum_me1000}
\end{figure}

\begin{figure}
\includegraphics[width=0.47\textwidth]{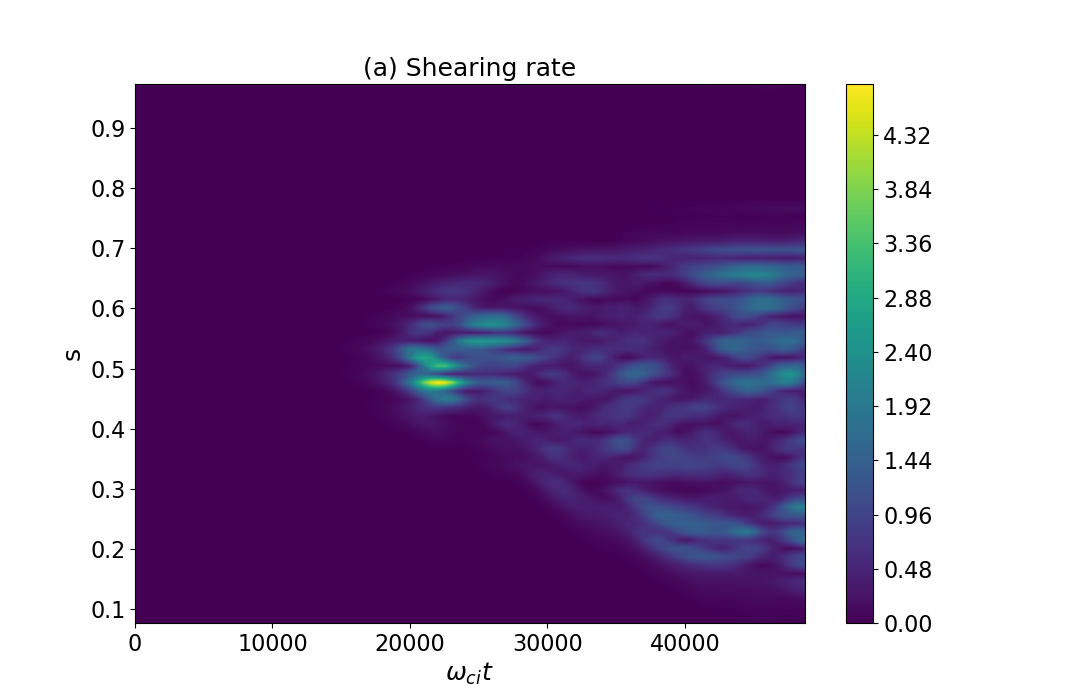} %{{./graph/BAE_b0.002_me1000/shearing_rate.png}}
\includegraphics[width=0.47\textwidth]{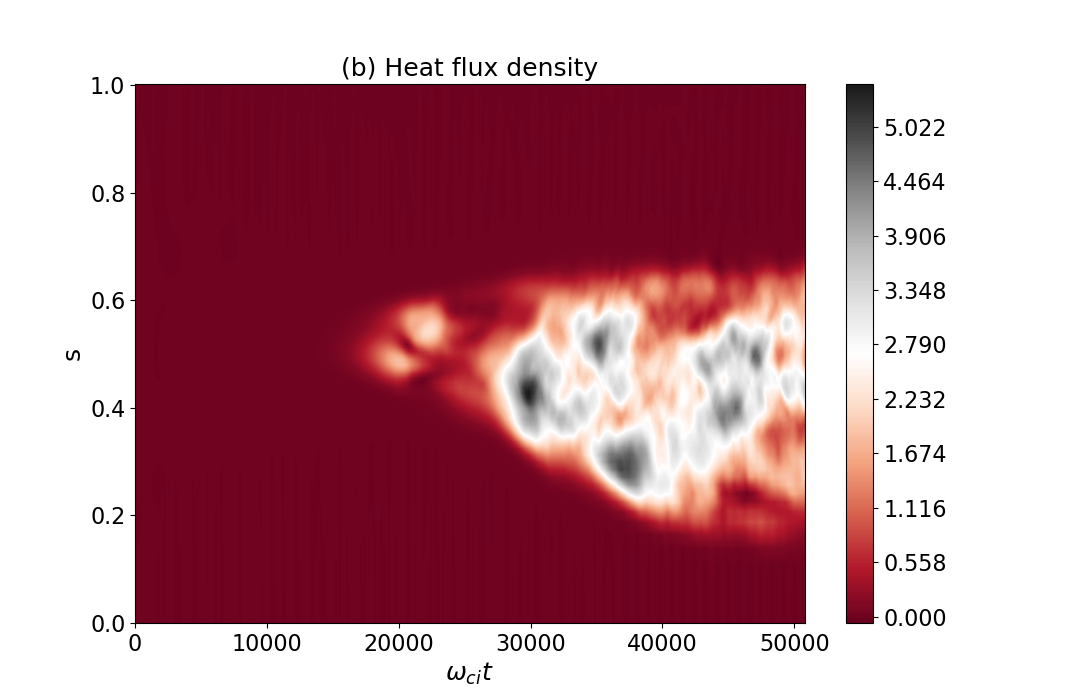} %{{./graph/BAE_b0.002_me1000/eflux_st.png}}
\caption{\Brunner{
(a) The zonal-flow shearing rate normalized to $\gamma_{lin}$ and (b) the radial heat flux evolution plotted for $\beta = 0.8\%$, $L_x = 350$, and $m_i/m_e = 1000$. One sees how the electromagnetic turbulence spreads in the regions with large temperature gradients.
}}
\label{shearing_me1000}
\end{figure}

We now turn our attention to increasing the mass ratio from $m_i/m_e = 200$ used so far in this paper, except in Fig.~\ref{gamma_eflux}(b), to higher values. An obvious difficulty of this parameter adjustment is the need to decrease the time step while keeping the simulation length in the physical time approximately the same, leading to a longer duration of the simulation. For the simulations shown in Fig.~\ref{gamma_eflux}(b), we have observed that the scaling of the time step with $\sqrt{m_i/m_e}$ is sufficient. In Fig.~\ref{hflux_me1000}, the heat flux and the signal-to-noise ratio are shown for $\beta = 0.8\%$, $L_x = 350$, and the mass ratio $m_i/m_e = 1000$. One sees that the noise is at the level of $1\%$ in this case. The electrostatic potential evolution is shown for this case in Fig.~\ref{potsc_me1000}. In this case, we can see the BAE-type structure developing similar to Figs.~\ref{potsc_KBM} and \ref{potsc_lx480}. Also in the evolution of the toroidal spectrum, shown in Fig.~\ref{spectrum_me1000}, one can see how the initial peak at $20 < n < 30$ is replaced by the distinct BAE component at $n = 5$ and its nonlinear harmonics at $n = 10$ and $n = 15$.
\begin{figure}
\includegraphics[width=0.47\textwidth]{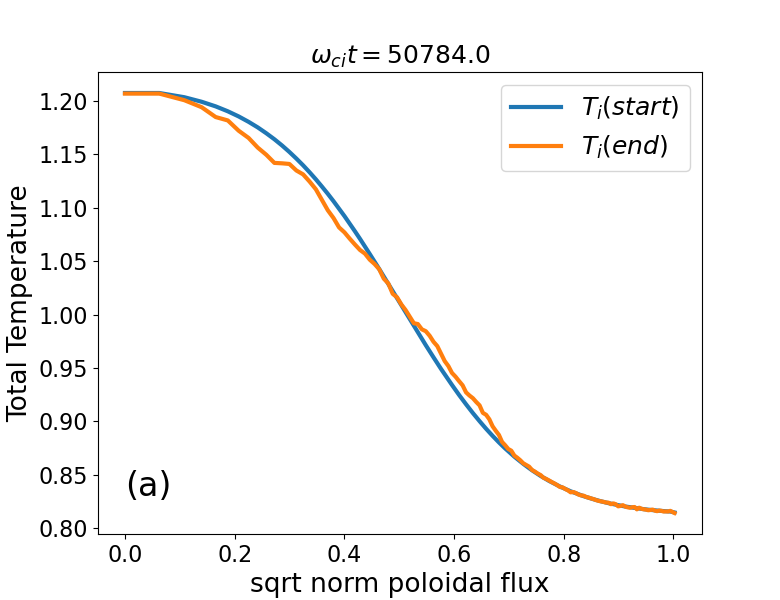} %{{./graph/BAE_b0.002_me1000/temp_relaxed.png}}
\includegraphics[width=0.47\textwidth]{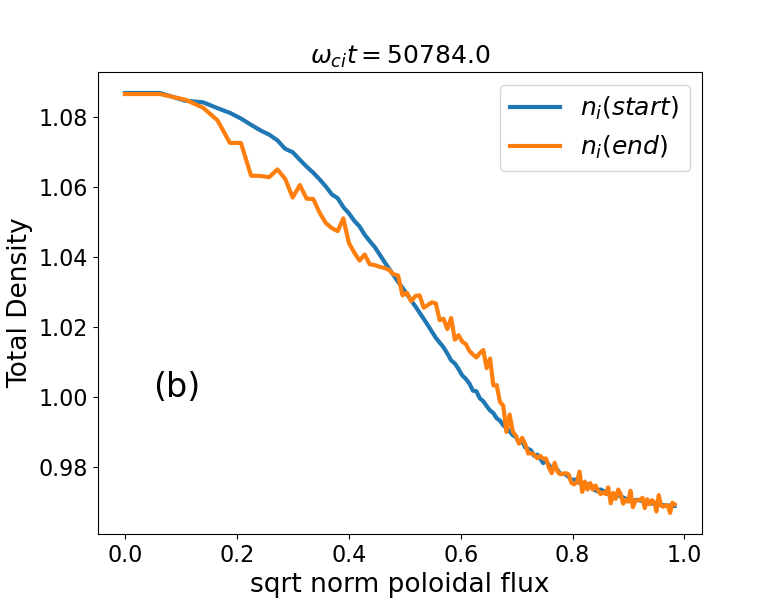} %{{./graph/BAE_b0.002_me1000/dens_relaxed.png}}
\caption{\Brunner{
(a) Temperature profile relaxation and (b) Density profile relaxation for $\beta = 0.8\%$, $L_x = 350$, and the mass ratio $m_i/m_e = 1000$.
}}
\label{temp_relaxed_me1000}
\end{figure}
In Fig.~\ref{temp_relaxed_me1000}, the temperature and the density profile relaxation, computed using Eqs.~(\ref{temp_def}) and (\ref{dens_def}), are shown for $\beta = 0.8\%$, $L_x = 350$, and $m_i/m_e = 1000$. One sees that the relaxation is not very strong. The normalized shearing rate and the radial heat flux evolution are plotted in Fig.~\ref{shearing_me1000} showing the turbulence spreading in the regions of the large temperature gradients.
%
%%%%%%%%%%%%%%%%%%%%%%%%%%%%%%%%%%%%%%%%%%%%%%%%%%%%%%%%%
%
\subsection{Simulations in shaped plasmas}
\begin{figure}
\includegraphics[width=0.47\textwidth]{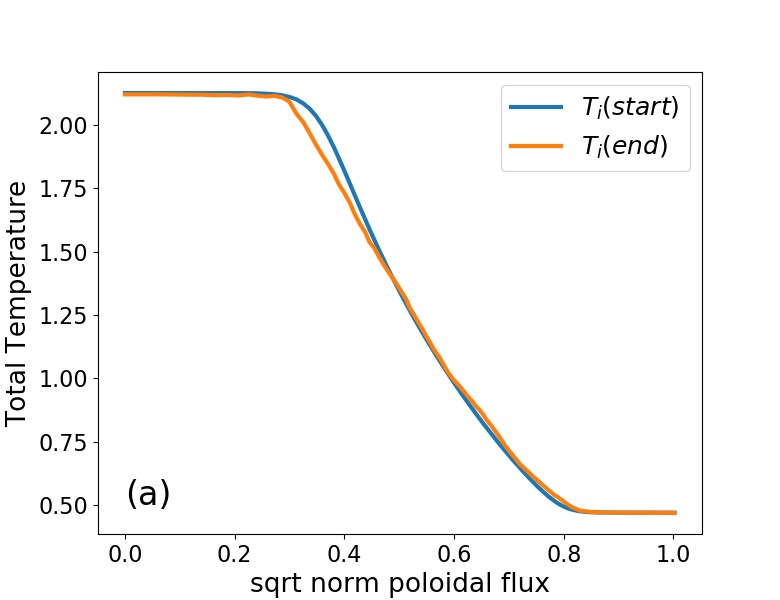} %{{./graph/temp_relaxed_iter.png}}
\includegraphics[width=0.47\textwidth]{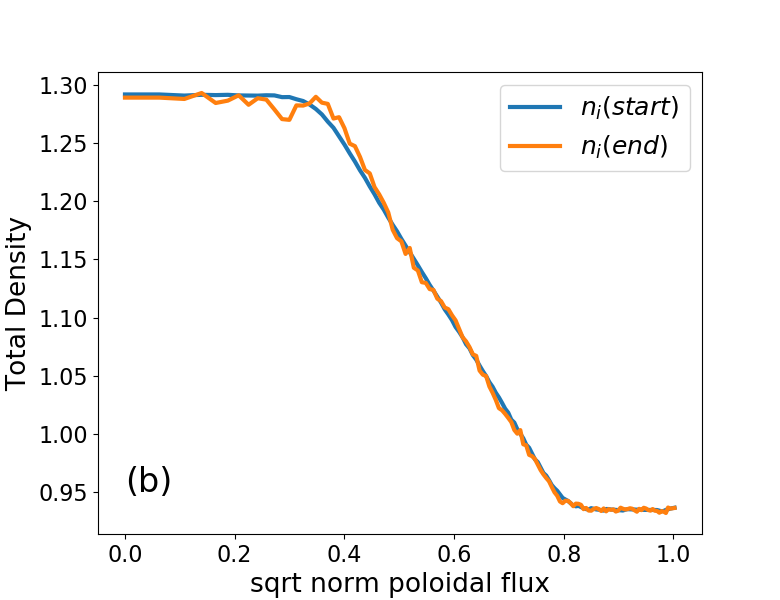} %{{./graph/dens_relaxed_iter.png}}
\caption{The total perturbed (a) temperature and (b) density profiles in the down-scaled ITER simulation (ITG-unstable). The profiles relax nonlinearly.}
\label{b0.00025_iter_profs}
\end{figure}

\begin{figure}
\includegraphics[width=0.47\textwidth]{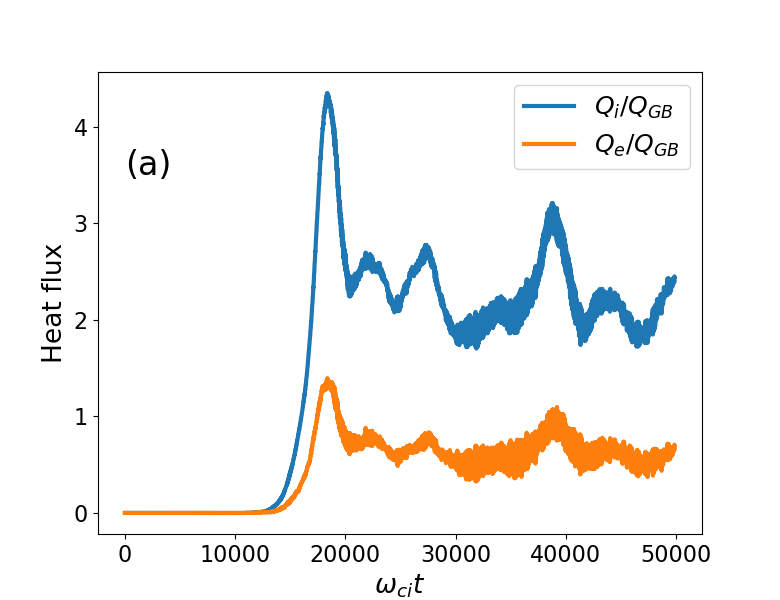} %{{./graph/eflux_t_iter.png}}
\includegraphics[width=0.47\textwidth]{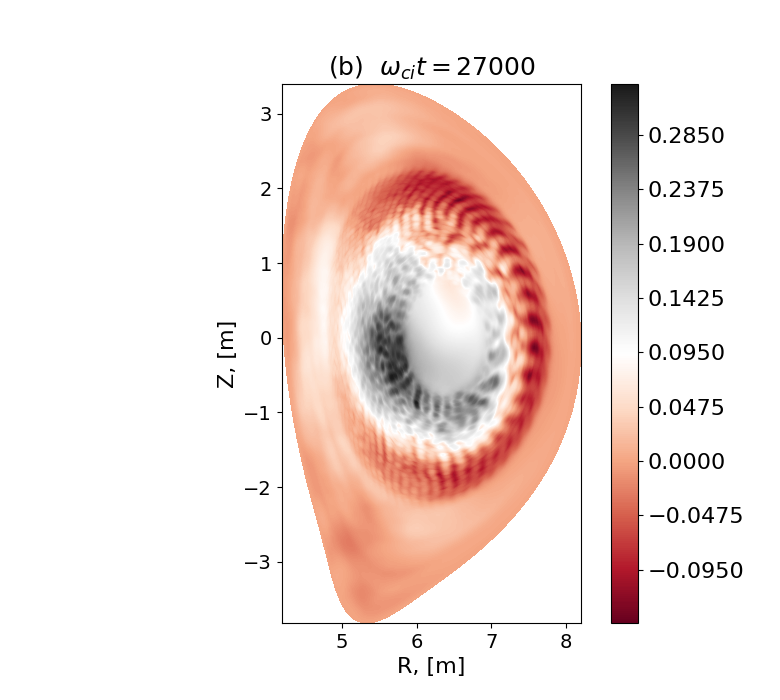} %{{./graph/iter_rz.png}}
\caption{(a) The averaged radial heat flux density and (b) the electrostatic potential in the poloidal cross-section for the ITER plasma scaled to the parameters of Ref.~\cite{Biancalani}. One sees that the turbulence reaches a saturated state.}
\label{b0.00025_iter}
\end{figure}

In this subsection, we attempt to relax the limitation of the simplified tokamak geometry considered in Ref.~\cite{Biancalani}. The main motivation for this attempt is the observation that the cancellation problem may become more complicated in shaped plasmas \cite{Mishchenko_shaping}. First, we follow Ref.~\cite{Hayward} and consider the ITER geometry with the appropriate magnetic field structure, aspect ratio, and the safety factor profile. This configuration is, however, scaled down to the parameters used in the majority of the runs of Ref.~\cite{Biancalani}, meaning small $L_x = 350$, $\beta = 0.001$, and $m_i/m_e = 200$. The temperature and density profiles employed in the simulation are shown in Fig.~\ref{b0.00025_iter_profs}. One sees that the profiles relax as a consequence of the turbulent dynamics.
In Fig.~\ref{b0.00025_iter}, the heat flux density and the structure of the electrostatic potential in the poloidal cross-section of this down-scaled ITER are shown. One sees that the turbulence is in the saturated state with the mode structure showing turbulent eddies broken by the zonal flows, qualitatively similar to the large-aspect-ratio tokamak, shown in  Fig.~\ref{potsc_ITG}(a).
\Sanchez{
For a long-time numerical stability, it was necessary to increase the width of the diagonal poloidal filter using 21 poloidal modes in the filter (instead of the 11 modes used before). As a consequence, the noise level is about $20\%$ in the nonlinear stage for the marker resolution available in the simulation, although it does not collapse into a numerical instability for quite a long time. A substantial increase in the computational effort will be needed in the future.
}
%Increasing the plasma beta, the mass ratio, and the machine size of the down-scaled ITER configuration to more realistic values will be addressed in the future numerical experiments.
%
%%%%%%%%%%%%%%%%%%%%%%%%%%%%%%%%%%%%%%%%%%%%%%%%%%%%%%%%%
%
%\subsection{Simulations in stellarator geometry}

\begin{figure}
\includegraphics[width=0.47\textwidth]{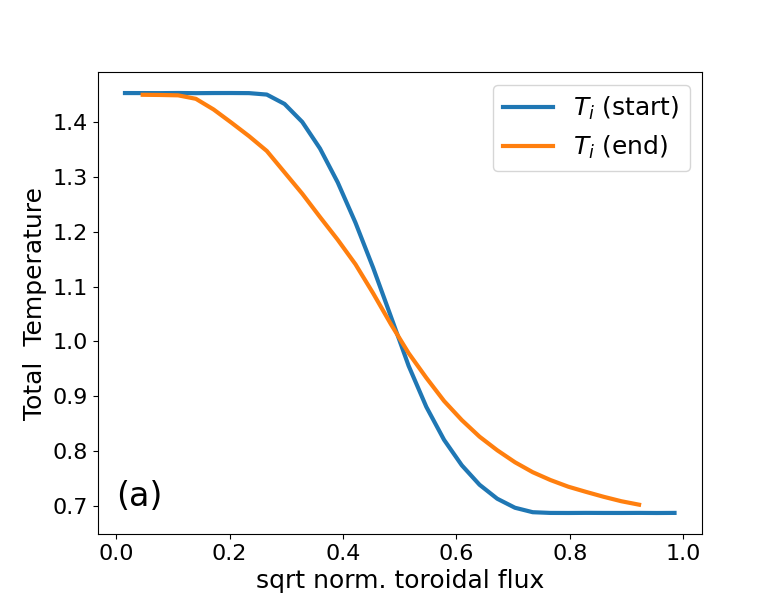} %{{./graph/temp_relax_w7x.png}}
\includegraphics[width=0.48\textwidth]{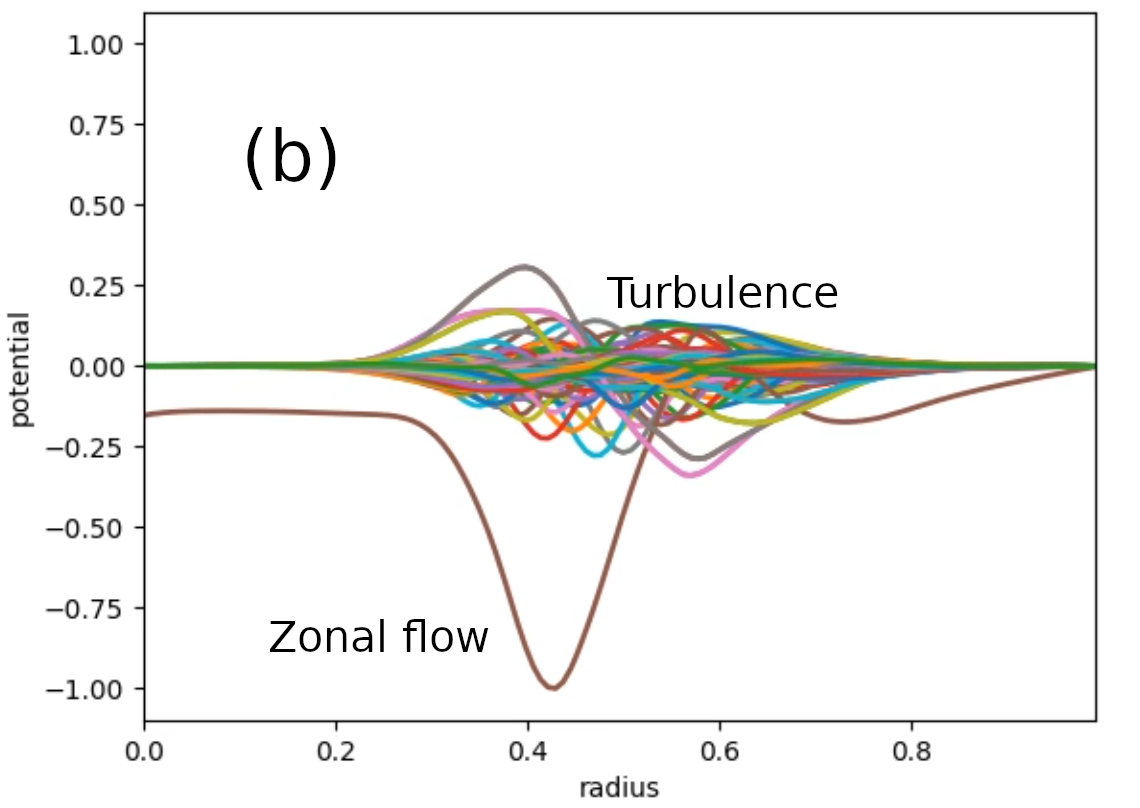} %{{./graph/dpot_sm_w7x.png}}
\caption{(a) The total perturbed temperature and (b) the poloidal Fourier harmonics of the electrostatic potential at the toroidal angle $\varphi=0$ for the down-scaled W7-X plasma. The turbulence generates zonal flows and enters the saturation phase. \Sanchez{One sees that the location of the turbulence and the zonal flow are correlated.}}
\label{turb_W7X}
\end{figure}

\begin{figure}
\includegraphics[width=0.47\textwidth]{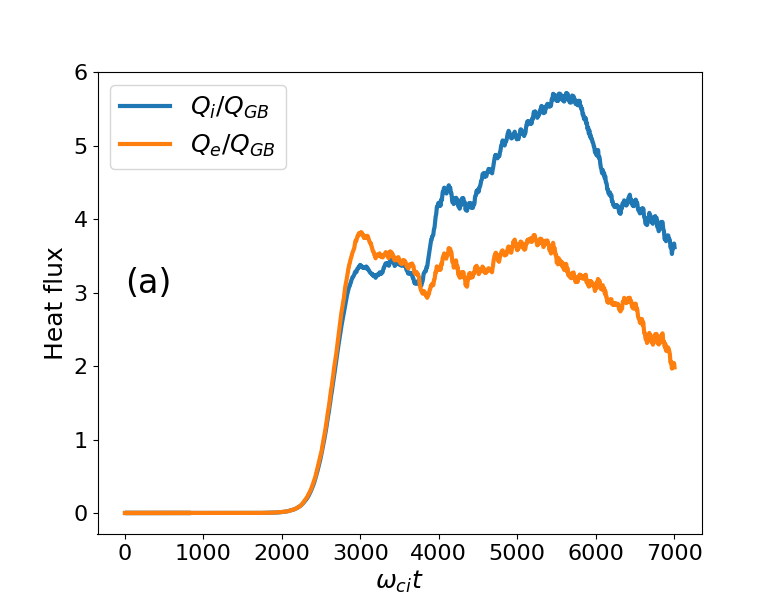} %{{./graph/eflux_t_w7xLx200.png}}
\includegraphics[width=0.47\textwidth]{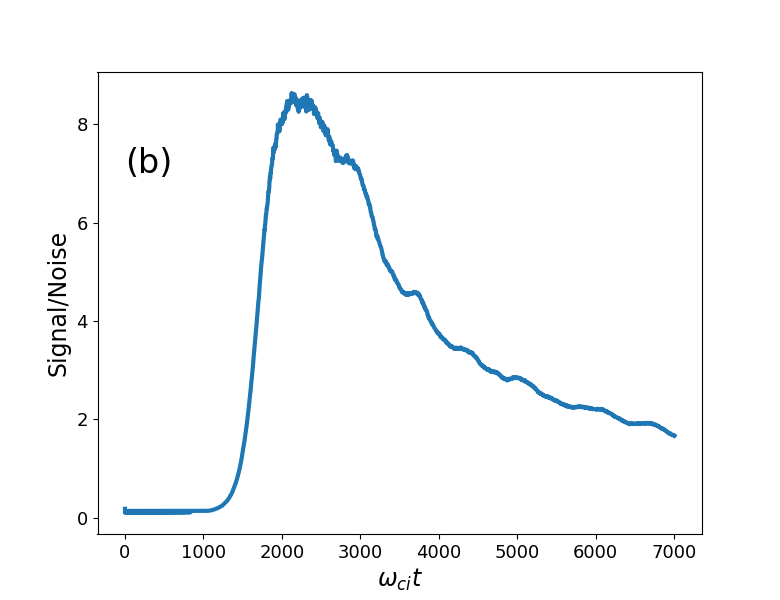} %{{./graph/SNR_t-w7x.png}}
\caption{(a) The radial averaged heat flux density and (b) the signal-to-noise ratio for $L_x = 200$, $m_i/m_e = 200$, and $\beta = 0.14\%$ in W7-X geometry. The gyro-Bohm units are used for the heat flux. \Sanchez{The heat flux decay in the final stage of the simulation is probably caused by the noise. The electrostatic potential shown in Fig.~21(b) corresponds roughly to $\omega_{ci} t = 3000$.}
}
\label{turb_W7X_flux}
\end{figure}

Finally, we relax the limitation of the axisymmetric ambient magnetic field and consider the stellarator geometry of Wendelstein 7-X (W7-X).
\Sanchez{
In contrast to all previous simulations, we use EUTERPE code and the SKL Irene supercomputer (TGCC) for the stellarator runs.}
As in the ITER case considered above, we down-scale the real W7-X to the parameters similar to those used in Ref.~\cite{Biancalani}: the number of ion gyro-radii in the simulation box $L_x = 200$, $\beta = 0.14\%$, and $m_i/m_e = 200$.
The Krook operator with the relaxation rate $\gamma_{Krook}/\omega_{ci} = 10^{-5}$ is used for the noise control. The diagonal filter is applied with the half-width $\Delta m = 17$. The simulation is initialized with a random noise.
The flat density profiles and the ITG-unstable temperature profiles are chosen for both the ions and the electrons. The resulting evolution of the ion temperature profile and the mode structure in the saturation phase are shown in Fig.~\ref{turb_W7X}. One sees that the turbulence is located in the center of the simulation domain where the temperature gradient has its maximum. The zonal flow generated by the turbulence has its maximum, as expected, in the region of the strong ITG activity. As a consequence, the weakly-electromagnetic turbulence enters a saturation regime. The temperature gradient is nonlinearly relaxed by the turbulence.
\Sanchez{The average heat flux evolution and the signal-to-noise ratio are plotted in Fig.~\ref{turb_W7X_flux}. One sees that the noise becomes quite large in the late phase of the simulation although it does not dominate the signal. Increasing the marker resolution would mitigate this issue but it would be very time-consuming (120 hours on $1920$ CPU cores were needed for the run shown in Fig.~\ref{turb_W7X_flux}). Still, it is encouraging that the electromagnetic PIC simulation of this type in the global stellarator geometry has run for quite a long time without a collapse. In future, a careful study of the noise correction and its effect on the stellarator zonal flow and turbulence, similar to Ref.~\cite{Sanchez_EUTERPE}, will be needed in the electromagnetic regime.}

%
%%%%%%%%%%%%%%%%%%%%%%%%%%%%%%%%%%%%%%%%%%%%%%%%%%%%%%%%%
%
\section{Conclusions}  \label{Conclusions}
In this paper, we have considered electromagnetic turbulence in tokamak and stellarator plasmas. The large-aspect-ratio tokamak and down-scaled ITER and W7-X geometries have been considered. In the large-aspect-ratio tokamak, we have shown that the extension of the results of Ref.~\cite{Biancalani} to larger plasma beta, larger machine size, and larger mass ratio is possible and can be affordable on existing computer systems \Sanchez{with a careful tradeoff between noise level and simulation cost.}

Simulations at a larger machine size appear to be the most challenging in terms of the computational cost since both the grid and the total number of markers have to be increased. As a consequence, a large number of nodes and GPUs is needed. Also, the usage of the GPU memory can become volatile and unstable when approaching the memory limit for any of the GPUs deployed.

For larger beta, including the perturbed magnetic flutter in Eq.~(\ref{bstar}) is mandatory to avoid numerical instabilities in the nonlinear phase. In terms of computational resources, the large-beta simulations are normally rather affordable since the grid and the marker resolution can often be the same as in the low-beta case. For larger beta, the nonlinear phase is dominated by a global mode with the periodicity of the BAE instability \cite{Biancalani_JPP}.
%The limit to increasing beta is set, for a given pressure gradient, by the numerical instability, probably related to the Shafranov-shift component of the pertubed magnetic potential, which appears in the nonlinear phase. This instability, however, disappears when the ambient pressure gradient is reduced. This problem can also be partially fixed increasing the numerical resolution.

Increasing the mass ratio requires longer simulations since the time step has to be reduced. This also makes the overall computation cost higher but there are less issues related to the number of the GPUs and the stability of the GPU memory providing that the grid resolution and the number of the markers do not change.

\Sanchez{
In the realistically-shaped down-scaled ITER geometry \cite{Hayward}, we have demonstrated numerically-stable saturated electromagnetic turbulence in the regime and for the parameters similar to Ref.~\cite{Biancalani}. However, the numerical noise was not small in this case for the marker resolution available in the simulation.
% Increasing the machine size, the plasma beta, and the mass ratio towards the realistic ITER values will be addressed in the future. Also, including the fast particles into the electromagnetic turbulence simulations at larger beta and considering configurations with MHD activities (such as the tearing modes), co-existing with the turbulence, will be important topics of future research.
%
The noise was also an issue in the stellarator (W7-X) electromagnetic turbulence in the nonlinear stage of the simulation. We will need a much larger marker resolution to suppress the noise in the realistic geometries. Presently, this requirement is in a conflict with the limitations imposed by the GPU memory. In future, substantial computational resources and an increased attention to the noise suppression techniques will be needed to simulate electromagnetic turbulence under realistic conditions with the gyrokinetic PIC codes.
}
%
%=================================================================
%
\section*{Acknowledgments}
This work has been carried out within the framework of the EUROfusion consortium and has received funding from the Euratom research and training programme 2014--2018 and 2019--2020 under grant agreement No 633053. The views and opinions expressed herein do not necessarily reflect those of the European Commission. This work was supported in part by the Swiss National Science Foundation.
We acknowledge PRACE for awarding us access to Marconi100 at CINECA, Italy, and to Joliot-Curie at GENCI@CEA, France.
Finally, we thank the organisers of the Spanish Fusion HPC Workshop and, in particular, Edilberto Sanchez, for inviting this paper.

%\section*{References}
%\bibliographystyle{unsrt} % or unsrt
%\bibliography{em_turb.bib}

\section*{References}
\bibliographystyle{unsrt} % or unsrt
%\bibliography{em_turb.bib}

%\begin{harvard}
%\item Hatzky2019
%\end{harvard}

\end{document}